\documentclass[12pt,preprint]{aastex}

\newcommand{\etal}{et al.}			

\newcommand{\kms}{km~s$^{-1}$}		
\newcounter{subfigure}

\def\H0{$H_0$~= 75 \kms\ Mpc$^{-1}$}

\def\ref{\par\noindent\hangindent 30pt}

\def\v16{$\Delta V_{1-6}$}
\def\lea{\mathrel{<\kern-1.0em\lower0.9ex\hbox{$\sim$}}}
\def\gea{\mathrel{>\kern-1.0em\lower0.9ex\hbox{$\sim$}}}

\begin{document}

\title{The Luminosity Function of Star Clusters in 20 Star-Forming Galaxies Based on  Hubble Legacy Archive Photometry\altaffilmark{1} }

\author{\sc Bradley C.\ Whitmore\altaffilmark{2}, 
Rupali Chandar\altaffilmark{3},
Ariel S.\ Bowers\altaffilmark{2,4},
Soeren Larsen\altaffilmark{5},
Kevin Lindsay\altaffilmark{2},
Asna Ansari\altaffilmark{2,6},
Jessica Evans\altaffilmark{7}
}

\affil{email: whitmore@stsci.edu, etc}

\altaffiltext{1}{Based on observations with the NASA/ESA {\it Hubble
Space Telescope}, obtained at the Space Telescope Science Institute,
which is operated by the Association of Universities for Research in
Astronomy, Inc. under NASA contract NAS5-26555. 
Also based on data obtained from the Hubble Legacy Archive, which is a
collaboration between the Space Telescope Science Institute
(STScI/NASA), the Space Telescope European Coordinating Facility
(ST-ECF/ESA) and the Canadian Astronomy Data Centre (CADC/NRC/CSA).
 Support for Program number 11781 was provided by NASA through a grant from the Space Telescope Science Institute.}
\altaffiltext{2}{Space Telescope Science Institute, 3700 San Martin
Drive, Baltimore, Maryland 21218, USA}
\altaffiltext{3}{University of  Toledo, Department of Physics \& Astronomy, 
Toledo, OH 43606, USA}
\altaffiltext{4}{Johns Hopkins University, Baltimore, MD, 21218}
\altaffiltext{5}{Department of Astrophysics/IMAPP, Radboud University Nijmegen, 
P. O. Box 9010, 6500 GL Nijmegen, The Netherlands}
\altaffiltext{6}{Northwestern University, Evanston, IL, 60201}
\altaffiltext{7}{Univeristy of Illinois, Champaign, IL, 61820}

\begin{abstract}
Luminosity functions have been determined for star cluster populations
in 20 nearby (4--30 Mpc), star-forming galaxies based on ACS source lists
generated by the Hubble Legacy Archive. These cluster catalogs provide
one of the largest sets of uniform, automatically-generated cluster
candidates available in the literature at present.  Comparisons are
made with other recently generated cluster catalogs demonstrating that
the HLA-generated catalogs are of similar quality, but in general do
not go as deep.  A typical cluster luminosity function can be
approximated by a power-law, $dN/dL \propto L^{\alpha}$, with an
average value for $\alpha$ of $-2.37$ and RMS scatter = 0.18 when using the  
F814W (``$I$'')~band.  A comparison of fitting results based on methods which use
binned and unbinned data shows good agreement, although there may be a
systematic tendency for the unbinned (maximum likelihood) method to
give slightly more negative values of $\alpha$ for galaxies with steeper
luminosity functions. We find that galaxies with high rates of star formation (or equivalently,
with the brightest or largest numbers of clusters) have a slight tendency to
have shallower values of $\alpha$. In
particular, the Antennae galaxy (NGC~4038/39), a merging system with a
relatively high star formation rate, has the second flattest
luminosity function in the sample.  A tentative correlation may also be present between 
Hubble Type  and values of $\alpha$, in the sense that later type galaxies (i.e., Sd and Sm)
appear to have flatter luminosity functions.  
Hence, while there do appear to be some weak correlations, the relative
similarity in the values of $\alpha$ for a  large number of
star-forming galaxies suggests that, to first order, the LFs are
fairly universal. We examine the bright end of the
luminosity functions and find evidence for a downturn, although it
only pertains to about 1\% of the clusters.  Our uniform database
results in a small scatter ($\approx$0.4 to 0.5~mag) in the correlation between
the magnitude of the brightest cluster ($M_\mathrm{brightest}$) and log of
the number of clusters brighter than $M_{I} = -9$ (log~N).  We also
examine the magnitude of the brightest cluster vs.\ log SFR for a
sample including both dwarfs galaxies and ULIRGS. This shows that the correlation
extends over roughly six orders of magnitudes but with scatter that is
larger than for our spiral sample, probably because of the high levels
of extinction in many of the LIRG galaxies.
\end{abstract}

\keywords{galaxies: star clusters---galaxies: interactions---galaxies:
individual (NGC~45 -- NGC~406 -- NGC~628 -- NGC~1300 -- NGC~1309 -- NGC~1313 -- NGC~1483 -- NGC~2397 -- NGC~3627 --  NGC~4038/39  -- NGC~4258 -- NGC~4394 -- NGC~4395 -- NGC~4736 -- NGC~5055 -- NGC~5236 -- NGC~5457 -- NGC~6217 -- NGC~6503 -- NGC~7793 )}

\section{Introduction}

Luminosity functions (LFs) provide a basic parameterization of the
star cluster population in galaxies. While the cluster mass function
is more fundamental, in many cases the multi-wavelength data necessary
to age-date the cluster population, and hence determine the cluster
masses, does not exist. In addition, the LF is directly observable and
does not require the use of a stellar population model that is
inherently uncertain.  To the degree that the star formation history
of different galaxies are similar, the luminosity function can serve
as an approximate proxy for the mass function. Some recent questions
being addressed using LFs include: 1)~how uniform are cluster
luminosity functions and what properties do they correlate with (e.g.,
with star formation history?), 2)~what is the shape of the LF and is
there evidence of a change of slope at either the faint or bright end,
3)~can the brightest cluster (hereafter $M_\mathrm{brightest}$) vs.\ log~N
(number of clusters brighter than $M_{I} = -9$) relation be
described as a single power law with a scatter which is only
statistical in nature?  Progress answering these questions is hampered
by non-uniformity in both the available observational datasets and the
criteria used to select clusters. In this paper we provide a large,
uniform (Hubble observations using the Advanced Camera for Surveys
[ACS]), and automatically-generated (Hubble Legacy Archive [HLA])
database to better address these and other related questions.

The current paper is part of a larger project that
aims to detect star clusters in several hundred galaxies using HST
imaging.  In this first paper, we outline the basic steps used in the
determination of luminosity functions for 20 of the galaxies with the
highest quality data.  We pay particular attention to the selection of the clusters, making
comparisons with other recently generated cluster catalogs in order to
estimate the degree to which selection effects can impact the results.  

This paper is organized as follows.  Section~2 describes the initial
source lists created by the Hubble Legacy Archive.  Section~3
describes some additional processing, and the method used to select
the clusters.  Section~4 presents the luminosity functions for the
target galaxies, including coadditions of galaxies with similar
properties (i.e., ``super galaxies''), and Section~5 discusses a
variety of correlations and their implications.  Finally, we summarize
our main results in Section~6.
The photometric catalogs used in this study are available at 
http://archive.stsci.edu/prepds/starclust-sfgal/.

\section{Initial Source Lists from the Hubble Legacy Archive (HLA) }

In recent years, star cluster luminosity functions have been determined for 
a relatively large number of individual galaxies using a wide range of 
observational setups and selection criteria. However, combining these
different studies in an attempt to determine general correlations
is problematic due to this diversity (see Portegies Zwart et~al.\ 2010 for a recent discussion). 
  
A primary goal of the current project is to select star clusters
in 20 nearby star-forming galaxies in a \textit{homogeneous} and \textit{objective} manner.
We only use observations taken with the 
ACS on the Hubble Space Telescope  and employ DAOPHOT-based source lists from the
Hubble Legacy Archive (HLA; available at http://hla.stsci.edu).

Below, we summarize the basic techniques used to develop the HLA
source lists, and refer interested readers to:
http://hla.stsci.edu/hla\_faq.html\#Source1
for more details.

When available, the HLA combines multiple images from different
filters in a visit into a single image to form a ``detection'' image
(also known as a ``total'' or ``white light'' image). This technique
allows detection of fainter sources with a wider range in colors and
ages than observations taken in a single filter, and provides a master
file with matched source positions in each filter.  A disadvantage is
that the completeness thresholds are not uniform for any given filter.

The HLA provides two source lists; one based on DAOPHOT (Stetson 
1987) for stars and compact objects and one based on SExtractor
(Bertin 1996) for extended objects. This paper is based on the DAOPHOT
HLA catalogs, since we are interested in objects that are slightly broader than the PSF
(i.e., compact clusters), for galaxies observed using  ACS.  Theses catalogs
use the routine DAOFIND on median-divided images (see Miller \etal\/
1999 for a discussion of this technique) to eliminate
common problems related to detecting objects in regions of bright and
variable local backgrounds, for example, in the high background areas
near the center of a spiral galaxy.

The measurements are made by performing circular aperture photometry
using the DAOPHOT/PHOT package/task in IRAF with a
3~pixel ($0.15''$) radius and the ABMAG photometric system. 
For consistency with most previous work in this field, 
we prefer to use the VEGAMAG system.
We therefore convert the photometry provided
by the HLA on the ABMAG system to the VEGAMAG system using Table~10 from
Sirianni \etal\/ (2005) for ACS.

\section{Additional Processing and Selection of the Clusters}

\subsection{Galaxy Sample}

In this paper, we select 20 star-forming galaxies with 
some of the highest quality ACS multi-wavelength observations available.
The galaxies and observations satisfy the following criteria:
(1)~distance modulus $m-M \lea 32$ (i.e., distance $\lea$ 30~Mpc)
(2)~galaxy type of Sa or later;
(3)~images taken with the Wide Field Camera of ACS  in the F814W ($\approx I$~band)
and at least one other broad-band optical filter, preferably a $V$~band filter;
(4)~source lists exist in the HLA as of October 2010, and reach to at
least $M_I \approx -9$.

The sample galaxies are listed in Table~1, along with some of their
basic properties. These include name, galaxy type, distance modulus,
foreground extinction, absolute magnitude M$_B$, proposal ID and
visit, and a list of the ACS filters available for each galaxy (note:
WFPC2 F336W observations are also included in italics as reference
information, but are not used in the current paper).  We also provide
an estimate of the star formation rate within the observed ACS field
for each galaxy, as described in Section~3.5.

Color images of the galaxies and the nine brightest clusters are shown in Figure~1.

\subsection{Cluster Selection}

The selection of star clusters from observations that contain
individual stars (both foreground and within the target galaxy),
background galaxies, clusters and associations, is one of the most important steps
for studies of extragalactic cluster systems.  In principle, an
objective, completely automated technique would be preferred, especially for
large datasets and samples of many galaxies, such as in the current study.  
However, a combination of potential problems, ranging from crowding to dimming of clusters
with age, make this approach problematic. For this reason, many
authors use both (or a combination of) automatic and manual techniques
(e.g., Chandar et~al.\ 2010, Bastian et~al.\ 2012, Johnson et~al.\ 2012). This is especially
appropriate for detailed studies of individual galaxies where
the goal is to push to the faintest levels.

For the present study, the approach is to examine cluster luminosity
functions, and in the future, mass functions, for a large homogenous
sample of galaxies. This  requires primarily an automatic
approach. One of the difficulties resulting from this approach is the
selection of faint clusters without including large numbers of
contaminants, such as close pairs of stars.  Because of this problem, the
limiting magnitudes from this study are brighter than are typically
found for detailed studies of cluster systems in individual galaxies.  The present paper
should therefore be viewed as being complementary to many of the
current papers which study the star clusters in nearby galaxies in more detail. Its
strength is that it is large, homogeneous, and objective; its weakness
is that it can not go as faint as more focussed studies of single galaxies.

While our general approach is to rely on automated selection for the
vast majority of the cluster candidates, a manual examination is made
for the brightest 10 clusters in each galaxy, since many of the
science goals of our project are focussed on these clusters (i.e., the
$M_\mathrm{brightest}$ vs.\ log SFR diagram and the possible turnover at the
bright end of the luminosity function). Automatic and manual
techniques have been compared in various studies (e.g., Chandar et~al.\ 
2010; Bastian et~al.\ 2012; Johnson et~al.\ 2012; Chandar et~al.\ 2014).  
While selection effects are always a concern, results
based on different cluster catalogs are usually quite similar
(e.g., Chandar et~al.\ 2010 find differences at approximately the 0.1
level when determining values of $\alpha$ using a wide variety of
selection criteria). We revisit this question in Section~3.3 with similar results.

Examples of common contaminants found in the catalogs when manually checking the 10 brightest clusters
are: 1)~bright (saturated) foreground stars, 2)~the nucleus of the galaxy, and 3)~the presence of bright background galaxies.

The primary diagnostic used for separating stars from clusters in an
image is whether the object has the same brightness profile as a star
(i.e., the point spread function).  In the current
paper we follow the general approach used in earlier papers (e.g., Whitmore et~al.\ 2010). 
We first select training sets of stars and obvious isolated clusters
in each target galaxy.  These training sets help us determine the
best range in values of the concentration index, $C$ (defined as the difference
in aperture magnitudes determined using a 3~pix and a 1~pix radius for the HLA)
for separating stars from clusters.  

In this study, as also found in Chandar et~al.\ (2010), a cut in $C$
alone is not sufficient to separate close pairs of individual stars in
the most crowded star-forming regions.  In these regions, the
concentration index of individual or close blends of stars can be
overestimated, thereby raising the $C$ value for non-clusters above
the star/cluster cutoff.  This problem is greatest in images of nearby
galaxies, where the DAOPHOT catalogs can contain large numbers of
individual stars detected within a single, highly resolved star
cluster.  We are able to eliminate most of these contaminants by
selecting the brightest object with a $C$ value above our cut and then
removing any fainter candidate sources within a radius of
$R_\mathrm{neighbor}$ (typically with values in the range 10--20 pixels---depending 
primarily on the distance of the galaxy).  The program used
to make this determination is called UNIQPOS.  While this ``neighbor
removal'' step occasionally removes another real cluster nearby, the
vast majority of the removed objects are close pairs of stars rather
than bona fide compact star clusters.  The final values of $C$ and
$R_\mathrm{neighbor}$ applied to the source lists in each galaxy are compiled in Table~2.

We determine the optimal depth of each catalog based on the quality of
the observations for each galaxy separately.  To accomplish this, we
trade depth, and hence size of the catalog, with the fraction  of contaminants,
which increases at fainter magnitudes. We aim for contamination rates less
than 10--20\%. A spot check described in Section~5.2 for NGC~1300
and NGC~5457 finds contamination rates in the range 5 to 12\%,
compatible with our goal.  The limiting magnitude used for our cluster
samples are primarily driven by the distance of the host galaxy, but
there are other factors, such as crowding, exposure time, dithering
strategy, etc.  that can also affect this limit.

We use a simple prescription to determine aperture corrections,
(which converts the fixed 3-pixel radius aperture magnitudes
from the HLA to total magnitudes) for our clusters.  In general, we determine the mean
aperture correction from 3 to 10 pixels ($0.5\arcsec$) based on
DAOPHOT measurements for the ten brightest clusters. 
Clusters are removed from the determination if a manual examination
shows them to be in a very crowded region.  An additional correction of
0.10 mag is used to extrapolate from $0.5\arcsec$ to infinity, following
the procedure in Holtzman et~al.\ (1995).

A weakness of this approach is that by adopting a single value
of the aperture correction for all clusters, we overestimate the total
luminosity of very compact clusters and underestimate the total
luminosity of more extended clusters.  Hence, another method that
could be used would be to apply aperture corrections to each object
based on its measured value of $C$. In Chandar et~al.\ (2010), we
compared the luminosity function of clusters in M83 that result when
an average correction is applied to each cluster, versus if a
size-dependent aperture correction is applied to each cluster.  We
find that the method used to determine aperture corrections affect the measurement of
$\alpha$ at only the $\approx$0.01 mag level in this particular case.  

In the current paper, we also investigate the effect on our results if
we use a larger aperture (i.e., a 7~pixel radius with a background sky
annulus from 15 to 20 pixels), minimizing the need for an aperture
correction.  This captures a much larger fraction of the total light
from a cluster, at the expense of occasionally including nearby
objects in the aperture. Typical differences between the 3 and 7~pixel
radii are $\approx$0.7~mag, with scatter of about 0.4~mag. We use an
upper limit of 1.5~mag for the difference to guard against outliers
(i.e., bright nearby stars within 7~pixels). These are relatively
rare, but it is important to remove them since they can cause very
large apparent aperture corrections (i.e., up to 4~mag) in some cases.
As will be discussed in Sections~5.1 and 5.3, it is reassuring to find
that the difference between the use of 3 and 7~pixel apertures
introduces relatively minor effects on our results.

Finally, we correct each cluster for foreground extinction (listed in
Table~1).  We do not, however, correct the luminosities for any
extinction within the host galaxy itself.  This would only be possible
for a few galaxies in the current dataset due to the lack of $U$~band
photometry necessary to determine age, mass, and extinction for
individual clusters (e.g., see Chandar et~al.\ 2010).  However, we note
that in Section~5.1 we find that the RMS scatter in the
$M_\mathrm{brightest}$ vs.\  log~N and $M_\mathrm{brightest}$ vs.\ log SFR relations
is only about 0.4~mag, leaving very little room for large values of
extinction. Hence extinction is not likely to significantly affect our
results. This is not the case for all galaxies, however. For example,
in Section~5.3 it is shown that ULIRG galaxies can have much larger
values of extinction, dramatically increasing the scatter in the
$M_\mathrm{brightest}$ vs.\  log~N diagram.

In Table~2 we list the brightness limits adopted for each cluster catalog, the values of $C$
and $R_\mathrm{neighbor}$ used to construct the catalogs, magnitude of the
brightest cluster in the $I$~band $M_\mathrm{brightest}$, and log N (i.e., log
of the total number of clusters brighter than $M_I=-9$).

\subsection{Comparison with Bastian et al.\ (2012) and Chandar et al.\ (2014) Cluster Catalogs}

Recently, studies by Bastian et al.\ (2012) and Chandar et~al.\ (2014) have 
compared cluster catalogs in the galaxy M83, and have
discussed how the selection of clusters can
affect the resulting conclusions. In this section, we revisit this
issue by comparing the HLA-generated catalogs used in the present
study with the catalogs from these two studies in M83. 

As described in more detail by Chandar et~al.\ (2014), a comparison
between three catalogs (Chandar-automatic, Chandar-manual,
Bastian-hybrid) shows that about 70\% of the clusters are in common when
comparing any two of the three catalog. The biggest area of
disagreement is for clusters with ages less than 10~Myr, with Bastian et~al.\ 
including only relatively symmetric clusters while Chandar et al. also
include slightly more diffuse, asymmetric clusters. The top panel in Figure~2
shows a color image of a region in M83 (see Chandar et~al.\ 2014 to see
the location of this region  within  M83). We invite the reader to make their own
estimates of the likely clusters based on this figure before comparing
with the cluster catalogs below. The reader might also want to go to
the HLA (http://hla.stsci.edu) (image =
HST\_9774\_0f\_ACS\_WFC\_F814W\_F555W\_F435W)
to make the comparison using the interactive display, since the 
combination of a color image, contrast control, and ability to zoom can be very helpful.

The middle panel in Figure~2 shows a comparison of the HLA-generated
cluster catalog used in the present study (large red circles) with the
other three catalogs (Chandar-automatic in blue, Chandar-manual in
yellow, Bastian-hybrid in purple---see Chandar et al. 2014 for 
a detailed description). The primary difference
between the HLA-generated catalog and the other more ``specialized''
catalogs is that it does not go as deep.  This is because the number
of contaminants grows rapidly for magnitudes fainter than $M_I = -8$, which
is therefore used as the cutoff for the HLA-generated catalog for M83. The
more specialized catalogs can go deeper since either a manual step is
included which guards against most contaminants, or in the case of the
Chandar-automatic catalog, the parameters can be better tuned for a particular dataset.

The bottom panel of Figure~2 shows the results when comparable
magnitude thresholds are used. The agreement is relatively good, with
an average of 71\% of the clusters in common, based on an average of
all 12 pairwise comparisons between the four catalogs. More
specifically, the average value is $75 \pm 8$\% when the
HLA-generated catalog is included in the comparison and $68 \pm 5$\%
when it is not (i.e., they are the same within the errors).  Similar results are 
found for two other fields (one in a denser and one in a sparser region). These
numbers are similar to those found in the comparisons reported in
Bastian et~al.\ (2012) and Chandar et~al.\ (2014). We conclude that the
HLA-generated catalogs are of comparable quality, down to a
threshold of $M_{I} = -8$.

A few comments on specific objects provides more insight into the
relative strengths and weaknesses of each catalog. The automatic
catalogs have more examples of close pairs of stars that are
incorrectly selected as  clusters (e.g., the red circle just above
object 5652 and the upper right most blue circle). This is the primary
type of contaminant included in the  automatic catalogs at fainter magnitude, as
discussed above.  Another important difference is that the Bastian
catalog has fewer objects in crowded regions with high backgrounds
(e.g., 5652 and 5715 are not included in the Bastian catalog).  This
is the main reason for a discrepancy between the Chandar and Bastian
catalog for clusters with ages $<$10~Myr (i.e., since young clusters
tend to be found in crowded regions), as discussed in both papers.

\subsection{Differentiating Young and Old Clusters}

Because there is color information for each of our
target galaxies, we can make a general assessment about the ages of
star clusters that dominate each galaxy sample, at least at the bright end 
where we have examined each cluster manually.  This is important
because the luminosity functions of ancient globular clusters are
quite different from those of compact young star clusters.  In particular, 
luminosity functions of young clusters tend to be well approximated by a 
power-law, while the luminosity (and mass) functions of old GCs have a 
pronounced deficit of objects below $M_I\approx-8.5$ ($\approx 2\times 10^5 M_\odot$) 
and appear peaked when plotted in logarithmic bins.  

Table~2 contains the total number of clusters brighter than $M_I=-9$ 
for each sample, as well as the $I$~band magnitude of the brightest cluster.  
Table~2 also includes an estimate of the number of the brightest 10
clusters that appear to be red, ancient globular clusters, rather than
young, blue clusters.  We make this manual assessment based on the HLA color
image.  Ancient globular clusters can generally be distinguished by the
fact that they;  1)~have redder colors because they contain only low mass
stars, 2)~they have small variations in their pixel-to-pixel flux
compared with younger clusters (e.g., Whitmore et~al.\ 2011), 3)~they tend
to have larger half-light radii than many of the very young clusters
(e.g., Bastian et~al.\ 2012), and 4)~they are generally found in uncrowded
regions away from regions of recent star formation.  One potential
complicating factor is that a young cluster behind a dust lane can
appear to have colors similar to an ancient globular clusters. However,
these can generally be identified by a manual examination (i.e., no
obvious dust lanes in the immediate area around the cluster).  We will
present a separate analysis of the old globular cluster systems in
these galaxies in a future paper (Bowers et~al.\ 2014). In Section~5.1
we find no statistically significant differences in the luminosity functions 
for galaxies with significant ``pollution'' by old globular clusters.

\subsection{Star Formation Rates}

One of the goals of this study is to examine the $M_\mathrm{brightest}$
vs.\ log~N relation, and the related M$_\mathrm{brightest}$ vs.\ log SFR relation. In
this section we describe how we measure the SFR for our galaxies.

We derived star formation rates for each field using GALEX far-ultraviolet
images and the calibration of SFR vs.\ UV continuum flux in
Kennicutt (1998). The calibration assumes a constant star formation
rate over time scales of $\sim$$10^8$ years, as is probably appropriate for most
of our targets, and is normalised to a Salpeter IMF with a minimum mass of $0.1\,M_\odot$.
The SFR and FUV continuum luminosity (here expressed as $L_\lambda$) are related as
\begin{equation}
  \mathrm{SFR} \left[M_\odot \, \mathrm{yr}^{-1}\right] = 1.07 \times 10^{-40} L_\lambda \left[\mathrm{erg\,s}^{-1} \, \mathrm{\AA}^{-1}\right]
\end{equation}
We obtained the FUV flux densities from the GALEV FUV (1500~\AA) images via the relation given on the 
GALEX web page\footnote{http://galexgi.gsfc.nasa.gov/docs/galex/FAQ/counts\_background.html} 
(see also Morrissey et~al.\ 2007):
\begin{equation}
  F_\lambda \left[\mathrm{erg\,s}^{-1} \, \mathrm{cm}^{-2} \, \mathrm{\AA}^{-1}\right] = {1.40\times10^{-15}} \, \mathrm{CPS}
\end{equation}
where CPS are the counts-per-second measured in the GALEX frames. The flux densities were converted 
to luminosities using the distances and foreground extinctions  in Table~1. The FUV extinction was computed 
assuming the relation between $E(B-V)$ and $A_\mathrm{FUV}$ in Hunter et~al.\ (2010), 
$A_\mathrm{FUV} = 8.24 \, E(B-V)$.

We note that the star formation rates were computed for the sections of the galaxies covered 
by the ACS/WFC footprints rather than for the whole galaxies. In practice, this was done by transforming 
the ACS images to the GALEX frames with the SREGISTER task in the IMAGES.IMMATCH package in 
IRAF and then masking out the regions of the GALEX frames that were not covered by the ACS images.  
These star formation rates are compiled in Table~1.

\section{Star Cluster Luminosity Functions}

In this section, we present luminosity functions for star clusters in
our program galaxies.  Previous works have found that luminosity
functions of star clusters can be described, at least approximately,
by a power law, $\phi(L) \propto L^{\alpha}$.  We first discuss three
different techniques we use to determine $\alpha$ in Section~4.1.
Results using different selection criteria for a single galaxy (M83)
are discussed in Section~4.2, followed by a discussion of the entire
galaxy sample in Section~4.3.  A comparison between our predicted
uncertainties, and empirical estimates of uncertainties (e.g., by
comparing different parts of the same galaxy) is made in Section~4.4.

\subsection{Techniques for Determining \boldmath{$\alpha$}}

Three different methods are used in this paper for the determinations
of $\alpha$.  The first two methods are simple fits to a binned luminosity 
function, one with constant sized magnitude bins---$\alpha_\mathrm{const-mag}$ 
and the other with fixed number of clusters  per bin $\alpha_\mathrm{const-number}$. 
See Maiz-Apellaniz \& Ubeda (2005) for a discussion of the pros and cons 
of the two methods. 

The third method uses a maximum likelihood power-law to fit  the luminosity function.
This method is independent of any binning of the data. If the LF is assumed to be a 
power-law $dN/dL \propto L^\alpha$ for $L_{\rm min} \leq L \leq L_{\rm max}$, the 
likelihood $\mathcal{L}(L | \alpha)$ of observing a dataset $L = \{L_1, L_2,
\ldots, L_n\}$ for a given $\alpha$ is given by
\begin{equation}
 \mathcal{L}(L | \alpha) = \left(\frac{\alpha+1}{L_{\rm max}^{\alpha+1} - L_{\rm min}^{\alpha+1}}\right)^n \prod_{i=1}^{n} L_i^\alpha
\end{equation}
Numerically, it is more convenient to work with the logarithmic likelihood,
\begin{equation}
  \log \mathcal{L}(L | \alpha) = 
  n \log\left(\frac{\alpha+1}{L_{\rm max}^{\alpha+1} - L_{\rm min}^{\alpha+1}}\right) + \alpha \sum_{i=1}^{n} \log L_i 
  \label{eq:ml}
\end{equation}
The maximum likelihood estimate of the slope, $\alpha_{\rm best}$ is
then the value for which $\log\mathcal{L}(L|\alpha)$ is
maximized. Once $\alpha_{\rm best}$ has been found, we estimate the
1$-\sigma$ confidence interval $\alpha_{\rm best}\pm\sigma_\alpha$ as

\begin{equation}
  \frac{
    \int_{\alpha_{\rm best}-\sigma_\alpha}^{\alpha_{\rm best}+\sigma_\alpha} \mathcal{L}(L | \alpha) \, d\alpha
  }{
   \int_{-\infty}^{\infty} \mathcal{L}(L | \alpha) \, d\alpha
  } = 0.68
\end{equation}

The value of $\alpha$ determined using this method will be denoted $\alpha_\mathrm{max-likelihood}$.

\subsection{The Effect of Changing Values of \boldmath{$C$} and \boldmath{$R_\mathrm{neighbor}$} on the Determination of \boldmath{$\alpha$} Using M83 as a Test Case}

In this section we assess the impact that our selection criteria have
on the resulting cluster luminosity functions.  In order to accomplish
this, we use the HLA catalog of field~1 of M83 (NGC~5236), and vary
the values of concentration index $C$ and $R_\mathrm{neighbor}$
(used by the UNIQPOS algorithm) discussed in Section~3.2.  For M83 we
found that the combination of $C \geq 1.15$ and
$R_\mathrm{neighbor}=20$~pixels gives the best results in terms of
separating the stars from clusters with the fewest number of contaminants
down to a limiting magnitude of $M_I=-8$.  In Figure~3, we present nine different
combinations, with $C$ values greater than 1.10, 1.15, and 1.20, and
$R_\mathrm{neighbor}=10$, 20, 30~pixels. These parameters span the range that produces
reasonable cluster catalogs in M83; going beyond these ranges results in catalogs 
that are clearly non-optimal based on  manual examinations.

The resulting values of $\alpha$ range from $-2.22$ (for $C = 1.20$,
$R_\mathrm{neighbor} = 30$) to $-2.37$ (for $C = 1.15$, $R_\mathrm{neighbor} = 10$). 
The mean for all nine determinations is $-2.30 \pm 0.05$. This
compares well with our preferred value of $-2.32 \pm 0.08$ based on the
fit with $C \geq 1.15$ and $R_\mathrm{neighbor}=20$ (i.e., the central panel).

A similar exercise has been performed using magnitude thresholds with
values $M_I = -7,$ $-7.5$, $-8.0$, $-8.5$, $-9.0$ magnitude (all with values
of $C \geq 1.15$ and $R_\mathrm{neighbor}=20$) in M83.  We find the values of
$\alpha$ range from $-2.38$ at $-7.5$ to $-2.28$ at $-8.5$, with a mean value
of $-2.34 \pm 0.04$, again in good agreement with our preferred value.
We also note that the fraction of contaminants (primarily pairs of stars---see 
Section~3.2) ranges from about 50\% at $M_I < -7$ mag to 20\% at 
$M_I < -8$ mag (similar to the results obtained in Section~5.2 for NGC~1300 
and NGC~5457) to about 10\% at $M_I < -9$ magnitude, based on visual examination.

We conclude from these exercises that the adopted parameters for our
best M83 cluster catalog (i.e., magnitude threshold $M_I < -8$
magnitude, $C \geq 1.15$ and $R_\mathrm{neighbor}=20$) are near optimal,
given our particular algorithms for selecting the clusters.  Reasonable
values of $C$ and $R_\mathrm{neighbor}$ result in changes to $\alpha$ at
about the 0.1 level. Care needs to be taken when determining the
magnitude limit to guard against the inclusion of large fractions of
contaminants when producing automatic catalogs.

\subsection{Cluster Luminosity Functions for the Entire Galaxy Sample}

Next, we examine the results for the ``best'' cluster catalogs for all the 
galaxies in the sample.  These catalogs were constructed using the $C$ 
and $R_\mathrm{neighbor}$ values reported in Table~2, down to the magnitude 
limit given in column~2 of the table.  We compile the best fit values of 
$\alpha$ for each of these catalogs in Table~3, for the max-likelihood and 
const-number fits.  The const-magnitude fits give similar results.

Figure~4 shows a comparison of the values of $\alpha_\mathrm{max-likelihood}$
with $\alpha_\mathrm{const-number}$. While the mean values are in reasonably
good agreement (i.e., $\Delta = -0.08 \pm 0.14$, and the agreement is
good in the range $\alpha$ = $-1.8$ to $-2.5$, there is a tendency for the
$\alpha_\mathrm{max-likelihood}$ values to be more negative than the
$\alpha_\mathrm{const-number}$ values for the steeper LFs. 

Overall, based on Figure~5, the cluster luminosity functions in all of our sample galaxies
appear to be reasonably well described by a single power-law. There is no obvious evidence 
for a break or change in slope at either the bright or faint end of these distributions (at least 
based on Figure~5), a topic that we will revisit in Sections~5.1 and~5.2.

Figure~5 shows the luminosity functions for each of our 20 program
galaxies.  We find values of the power-law indices, $\alpha_\mathrm{const-number}$, that 
range from approximately $-1.95$ (NGC~4395) to $-2.80$ (NGC~1483---however, 
note that the estimated error for this galaxy is 0.65).  This is a somewhat broader 
range, reaching slightly steeper values, than found in most studies to date. For example, 
Larsen (2002) finds a range from $-2$ to $-2.4$ for 6~spiral galaxies.  He also found a 
weak tendency for steeper slopes among fits to the brighter magnitudes. This might be 
part of the reason for steeper slopes in our results, since the limiting magnitude for our 
study was roughly a magnitude brighter than in the Larsen (2002) study. 

\subsection{Empirical Determination of Uncertainties }

The mean estimated uncertainties as returned by the software for the
various methods of determining values of $\alpha$ in Table~3 are 0.19
(maximum likelihood) and 0.21 (constant number). The empirical scatter
between determinations for each galaxy using the two methods is 0.14
magnitude, providing evidence that the estimated uncertainties are
fairly accurate, and may actually be slightly overestimated.

We can make additional empirical determinations of the true
uncertainties by comparing fields which have been observed in two
different fields for three of the galaxies (i.e., NGC~1300, NGC~5236,
and NGC~5457). There is no guarantee that the different fields have
the same values of $\alpha$, hence this should be considered an upper
limit. However, we note that in all three cases, the two fields are
roughly the same radii from the center of the galaxy (see Figure~1),
hence any trends caused by radial gradients would be minimized.  The
mean differences are 0.33 (maximum likelihood) and 0.19 (constant
number), in reasonable agreement with the numbers quoted in the first
paragraph.

We conclude that a typical value for the uncertainty in the
determination of $\alpha$ for our galaxies is in the range 0.15 to
0.20.  The overall scatter in the values of $\alpha$ for the entire
sample of galaxies are 0.27 for maximum likelihood and 0.18 for
constant number fits, comparable, or slightly larger than our
predicted uncertainties.  This suggests that much of the observed
spread in values of $\alpha$ for the entire sample is due to
statistical noise, rather than real differences between the galaxies.

Perhaps a better test of the degree to which the observed scatter is
real can be made by only including galaxies with error estimates $<$0.2.  
In this case we find mean values of $-2.39$ (maximum likelihood)
and $-2.35$ (constant number) with values for the observed scatter of 0.21
and 0.14.  The mean for the two methods yields $-2.37$ and a scatter of
0.18. We adopt these as our best-guess values for the remainder of the paper. 

The mean of the estimated uncertainties returned by the software, as
listed in Table~3, are 0.13 and 0.11 for this subset of high S/N galaxies.  
If accurate, this suggests that roughly 1/2 of the observed scatter is real 
(i.e., by subtracting the observed and estimated errors in quadrature). 
The weak correlations discussed in Section~5.1 and~5.2 provide 
further evidence that at least some of the observed scatter is real.

Table 3 includes a compilation of $\alpha$ estimates from other
studies.  The mean difference between our constant number bin measurements
of $\alpha$ and others is 0.12 with a scatter of 0.16. This again
supports our conclusion that the mean uncertainty in our estimates of
$\alpha$ is about 0.15 magnitude (for the higher S/N galaxies), and
also demonstrates the result that we tend to find somewhat steeper
values of $\alpha$ than many past studies.

\section{Results and Discussion}

\subsection{Searching for Correlations with the Slope of the
Luminosity Function}

Figure 6 shows a plot of $\alpha$ vs.\ the following parameters: a)~log
SFR, b)~distance modulus (DM), c)~$M_\mathrm{brightest}$, d)~log~N, e)~Hubble 
type (T), and f)~absolute B~magnitude ($M_B$). The $\alpha$ values used 
in Figure~6 are from the constant number fits. The maximum likelihood fits 
show similar correlations but with slightly more scatter, similar to the 
results discussed in Section~4.

The significance of each correlation (in units of $\sigma$, defined as
the slope of the correlation divided by the uncertainty in the slope
of the correlation) is provided in each panel. Typical differences
between the two methods are about 0.5$\sigma$.

The open circles in Figure~6 show the galaxies with very few clusters (i.e., 
log~N~$<1.2$ for the $M_I = -9$ limit: see Table~2) while the solid circles 
(i.e., the high S/N sample) show the 16 galaxies with log~N~$>1.2$.  The 
significance of the correlations using both the full and high S/N samples are 
included in Figure~6.

The strongest correlation (2.8$\sigma$) for the full sample is with
Hubble Type (T); with later-type galaxies having less negative values 
of $\alpha$ (i.e., a flatter LF).  It appears that the correlation may actually be slightly
U-shaped, with the earlier types $T < 4$ (i.e., Sbc and earlier
galaxies) also showing slightly lower values of $\alpha$. If we
restrict the range to $T > 4 $ (i.e., later than Sbc galaxies) the
correlations becomes even better (i.e., 4$\sigma$---see Figure~6).
We note, however, that removing NGC~4038/39 (the Antennae) reduces the 
correlations by about a factor of 2.

The next best set of correlations for the sample are with $M_\mathrm{brightest}$, log~N, 
and log~SFR, with a value of 2.0$\sigma$ being reached for the constant
number fits for $M_\mathrm{brightest}$. The $M_\mathrm{brightest}$ correlation for the high
S/N galaxies reaches a value of 4$\sigma$.  In all cases the trend is in the 
sense of the galaxies with more star formation (hence more clusters and a 
brighter values of $M_\mathrm{brightest}$ due to the size-of-sample effect) having 
lower values of $\alpha$. It is already well known that these three parameters 
correlate with each other well, as shown in Figure~7. This will be discussed in 
more detail in Section~5.3. Hence, these three correlations are probably 
manifestations of the same underlying correlation.  We note, however, that 
removing NGC~4038/39 (the Antennae) reduces the correlations by about a 
factor of 2 in these 3 cases.

We also note that the use of larger apertures (radii~= 7~pixels) in an
attempt to minimize the dependence on mean aperture corrections, as
discussed in Section~3.2, has only a minor effect on the results shown
in Figure~7. The slopes in the bottom two plots change by only~1 or 
2~percent (in both cases becoming slightly flatter) when using the
larger aperture measurements. The scatter and the statistical
significance improve, with values of RMS~= 0.44~mag and 10.6$\sigma$
for the $M_\mathrm{brightest}$ vs.\ log~N relation and RMS~= 0.36~mag and 
12.4$\sigma$ for the $M_\mathrm{brightest}$ vs.\ log~SFR relation. The scatter in
these diagrams is remarkably low!

The inclusion of low S/N data points may mask real correlations.  When
only the high S/N galaxies (solid points) are included in the fit we
find that the correlations improve in some cases (e.g., with
$M_\mathrm{brightest}$ and log~N), but get worse in others  (e.g., Hubble Type).

There may also be a weak correlation with the absolute B~magnitude of
the galaxy ($M_B$), though this becomes weaker for the higher S/N sample.

We note that there is little or no correlation with the distance
modulus (DM). This is reassuring and demonstrates that 
strong biases are not introduced by differences in spatial resolution.
Finally, there is no significant correlation between $\alpha$ and the
fraction of red (old) clusters, as discussed in 3.4 (not shown).

Figure 8 shows the correlations between Alpha ($\alpha$) and Hubble
type (T) using six different methods. This provides an indication of
how robust our results are.  The first three examples show the effect
that different methods of fitting the luminosity functions have on the
results (i.e., the maximum likelihood, constant number and constant
magnitude methods defined in Section~4.1). The other panels show the
effects of using larger apertures, and the brighter part of the LF. The 
high S/N dataset is used for all six panels along with a requirement 
that there be at least twenty clusters in the sample for each individual 
galaxy.  In all cases a weak 1.3 to 2.5$\sigma$ correlation is found. 
The constant bins fitting method gives the smallest scatter and has 
therefore been used as our primary method for most of this paper.

The bottom row of Figure~8 shows the results when using 7-pixel
photometry (using the constant magnitude method), as discussed in Section~3.2, 
in an attempt to minimize the dependence on aperture corrections. The same
general trend is seen although with larger scatter. The two panels labeled 
``bright'' show similar estimates but  with a faint magnitude limit which is 
1~magnitude brighter, in an attempt to determine whether there is a turndown 
at the bright end of the luminosity function. In both cases there is a marginal 
0.1~magnitude trend for the brighter fits to be steeper, suggesting a possible 
turndown. However, this represents less than a 1$\sigma$ difference in both 
cases. This point will be revisited in Section~5.2.

We finish by noting that while there do appear to be some weak correlations 
present, perhaps the main result is the apparent similarity of the values of 
$\alpha$ over a relatively wide range of galaxies. This suggests that to first 
order, the LFs are fairly universal, similar to results for mass functions (e.g., 
Whitmore et~al.\ 2007 and Fall and Chandar 2012---but see also Bastian 
et~al.\ 2012 and Chandar et~al.\ 2014).  However, with the availability
of a relatively large sample of galaxies (20), and using a more uniform set of 
data and analysis techniques than has been possible in the past, it appears 
that we are beginning to detect weak second-order effects that may be 
important for understanding the physics behind the demographics of 
cluster formation and destruction.

\subsection{The Luminosity Function of ``Super Galaxies''}

The cluster lists produced in this work can be combined in different
ways to improve statistics and to explore more subtle features in the
shape of the luminosity functions.  Here, we create cumulative
distibution functions (CDFs) for ``super galaxies'' by combining
galaxies as a function of: 1)~total brightness in the $B$~band (i.e.,
$M_B$), 2) log SFR, and 3) Hubble Type.  Three subdivisions are
included in each case (see notes to Table~3 for details).  The
Antennae galaxy, NGC~4038/39 has been left out of this exercise since
it is a merger remnant and hence may have different properties than
normal spiral galaxies (i.e., it has one of the flattest LFs).  In
support of this possibility, Randriamanakoto et~al.\ (2013) report
flatter LF's for their sample of LIRG galaxies, most of which are mergers.

The data used to define these CDFs have been fit in the same
way as the single galaxy luminosity functions; the results are
compiled in Table~3.  Figure~9 shows the CDFs for various
composite galaxies. Dashed lines show power law slopes of $-2.0$
(flattest), $-2.5$, and $-3.0$ (steepest) for reference.

We first note that most of the fits for $\alpha$ are between $-2.0$ and  $-3.0$
limits, with values around $-2.5$ being most typical. This is consistent
with values for $\alpha$ in Table~3 and Figure~6, as expected.  We
find that a few of the CDFs tend to steepen at the bright end,
providing tentative evidence of a turndown.  However, it should be
kept in mind that this represents a very small number of clusters
(i.e., generally the brightest 5--10 clusters out of several
thousand, i.e., less than 1\% of the data).

The graphical overemphasis on a relatively small number of bright clusters, 
which is implicit when showing cumulative distribution functions on a log scale, 
complicates comparisons with the regular power law fits of $\alpha$ listed in 
Table~3.  For example, based on the full CDF for the $0.3\,M_{\odot}$\,yr$^{-1}$ 
sample (i.e., the green line in the small inset of the middle panel of Figure~9) 
one might conclude that $\alpha \approx -3.0$. However, Table~3 gives a value 
of $-2.54$. This is because the vast majority of the clusters are faint (e.g., 77\% 
are fainter than $M_I = -10$), and the slope is flatter in this magnitude range, as
shown by a close look at the middle panel in Figure~9. The 50\% and 1\% 
points are included in Figure~9 to reinforce this point.

Of the nine composite galaxies shown in Figure~9, eight appear to have a 
downturn (i.e, the slope changes by more than 0.25 relative to the fiducial 
$-2$, $-2.5$, $-3$ slopes as one goes from the 50\% point to the 1\% point).  
The remaining case (Sc--Scd galaxies) has a relatively straight CDF in this 
range; no CDF has an upturn at the bright end. If we do a similar analysis 
of the individual galaxies (not shown), where low-number statistics are 
often an issue, we find that eight of the galaxies appear to have a downturn 
in their LFs, six appear relatively constant, and only two appear to have 
upturns.  Hence, there does appear to be some evidence for a downturn at 
the bright end of most of the CDFs. This is similar to the results discussed 
in Section~5.1 for the LFs.

A potential concern about the reality of the apparent downturn at
bright magnitudes is the fact that we manually inspected the top 10
clusters in each galaxy and removed contaminants (e.g., foreground stars,
galactic nuclei, background galaxies---see Section~3.2). This could, in 
principle, cause part of the downturn since we have not done the same 
for the fainter cluster candidates.

We can estimate the effect this might have by spot-checking a few of the 
galaxies at the fainter magnitudes to determine the fraction of
contaminants. Based on manual inspections of NGC~1300 and NGC~5457, we
find the following fractions of contaminants. For $M_I$ in the range $-10.5$
to $-10$ magnitude we find $12 \pm 6$\%, for $M_I$ from $-10.0$ to $-9.5$ 
magnitude we find $5 \pm 2$\%, and for $M_I$ in the range $-9.5$ to $-9$ 
magnitude we find $5 \pm 2$\% contaminants. We conclude that the contaminant
fractions are typically quite small, and are relatively constant as a function of $M_I$. 
Even if there were $\approx$20\% differences as a function of magnitude from 
brightest to faintest, this would only introduce changes in $\alpha$ which are
comparable to our statistical uncertainties (i.e., $\approx$0.1).  

We conclude that there is evidence that many of the galaxies have a small
downturn in their CDFs at bright magnitudes.  While the LF is clearly related to 
the MF,  there is not a simple 1:1 correspondence between the two  (e.g., see Fall 2006 
and Larsen 2009). Hence, we are not able to make a similar statement about the MF.
We leave a detailed investigation of this issue for the future. 

We now look at each of the three categories of composite galaxies in
turn, starting with the $M_B$ compilations on the top. We find
a possible weak trend for the CDF to be steeper for lower luminosity
galaxies at the 50\% point. However, the trend is essentially gone 
by $M_{I} = -11$. In addition, the trends in Figure~6 are not very 
significant, and are actually in the opposite direction. Hence we conclude 
that there is no clear correlation for $\alpha$ to change as a function of  $M_B$.

We next turn to the 3 SFR categories as shown in the middle panel of 
Figure~9. We find an apparent trend, with  a steep slope for the low SFR galaxies
($-2.88 \pm 0.22$), and  shallower slopes for the intermediate SFR category
($-2.56 \pm 0.09$) and  for the high SFR category ($-2.54 \pm 0.08$).
As discussed in the previous section, the same weak trend (and the associated 
$M_\mathrm{brightest}$ vs.\ log~N relationship) can be seen in Figures~6, 
providing additional evidence that this correlation is real.

The bottom panel shows the Hubble type composite galaxies. We have already 
seen in Figures~6 and~8 what appears to be a weak correlation  between values 
of $\alpha$ and~T.
In Figure~9, we see some evidence for the early type galaxies to be shallower, 
at least down to about $M_I = -10.5$ mag. However,  there is no clear trend for 
the Sd--Sm galaxies to be different than the Sc--Scd galaxies.  The values for 
$\alpha$ for the three Hubble type categories listed in Table~3 also give ambiguous 
results. Hence, the weak tendencies suggested by Figures~6 and~8 should be 
considered tentative. In particular, we note that the removal of the Antennae
would reduce the apparent correlation in Figure~6 by about a factor of two.

For this reason, the Antennae has been left out of the supergalaxy compilations 
and is shown separately in Figure~9. As expected, the CDF for the Antennae is 
significantly flatter than cumulative distributions for the other galaxies (note 
that due to incompleteness fainter than $M_{I} = -10$ mag, we have extrapolated 
down to $M_{I} = -9$ mag, as shown by the dotted line). This flatter slope is 
supported by Table~3 where the Antennae has the 2nd flattest slope of all 20 of the
program galaxies. Only NGC~4395, with large uncertainties due to very few clusters, 
has a flatter slope. 

\subsection{\boldmath{$M_\mathrm{brightest}$} vs.\ log N  and \boldmath{$M_\mathrm{brightest}$} vs.\  log SFR Relationships}

One of the goals of our study is to examine whether star formation in violent environments 
is fundamentally different than star formation in quiescent settings. Whitmore (2003) 
addressed this question by making a plot of the brightest cluster ($M_\mathrm{brightest}$) 
versus the log of the number of clusters brighter than $M_I = -9$ mag (log~N) in a galaxy.  
If there are two modes of star formation we might expect this distribution to be bimodal. 
However, the distribution is continuous, with a well defined correlation that can be 
explained by simple statistics; e.g., by drawing different size samples from the same 
$dN/dL\propto L^{-2}$ luminosity function.  This is often referred to as the 
``size-of-sample" effect.  Hence, it appears that the processes involved in the formation 
of star clusters are largely universal, and do not depend strongly on the environment.

Larsen (2002) confirmed this result in the form of the $M_\mathrm{brightest}$ vs.\  
log SFR relationship, and more recently, several authors have added additional data 
or developed more sophisticated analysis techniques (Whitmore, Chandar \& Fall 
2007; Bastian et~al.\ 2008; Larsen 2009;  Vavilkin 2011) .

Unfortunately, the data points used in these studies have several observational 
shortcomings. For example, they often use a very inhomogeneous data set (e.g., 
a mixture of ground-based and space-based observations, different wavelength 
ranges requiring extrapolations to the $V$~band, different analysis techniques 
since they originate from different papers, etc.). Using the results of the present, 
more homogeneous sample, we are now in a better position to re-examine this 
question.  

The top panel of Figure~10 shows the results for the 20 galaxies in our sample 
(note that there are actually 23 points since we have 3~galaxies with two 
separate fields of view). Two galaxies have completeness issues (i.e., NGC~6217 
and NGC~4038/39 are only complete to $M_I = -10$ mag). A correction has 
been made to these galaxies by measuring the fractional increase in the number of 
clusters brighter than $-10$ mag to the number brighter than $-9$ mag for 
NGC~1309, NGC~2397, and NGC 3627, and then applying this correction to 
the two incomplete galaxies.   The resulting correlation in Figure~10 is quite tight 
with a slope $-1.97 \pm 0.22$, RMS scatter~= 0.53 mag and 8.9$\sigma$ level of
significance. We note that this value is consistent with a value of $-1.82$ that 
would be predicted purely by statistics by a $L_\mathrm{brightest} \propto 
N^{-1/(1+\alpha)}$ dependence with a value of $\alpha = -2.37$ for the LFs in 
the sample (see Larsen 2009).


Hence, our uniform database does appear to result in smaller scatter in the 
$M_\mathrm{brightest}$ vs.\ log~N relationship than has been seen in past 
studies (e.g., Larsen 2002 finds RMS scatter~= 0.9 magnitude).

The bottom panel of Figure~10 shows the related $M_\mathrm{brightest}$ 
vs.\ log SFR diagram. Two additional datasets for LIRGs have been added 
to the diagram: 1)~Bastian et~al.\ 2008 (triangles) and 2)~Vavilkin 2011 
(asterisks). We note that part of the reason for the large difference between 
the SFR of our predominantly nearby spiral sample and the more distant LIRG
sample is that the field-of-view  for our sample of spiral galaxies typically
includes only a portion of the galaxy, while the entire galaxies are included in
the more distant LIRG samples.  Nevertheless, most of the difference in SFR 
is real, with the LIRG galaxies often having SFRs that are orders of 
magnitude larger than spiral galaxies. 

An advantage of the log SFR version of the relationship is that it also allows 
us to include  dwarf galaxies, which may not have a $M_I= -9$ mag cluster 
needed to define log~N.  Figure~10 shows that the correlation extends over 
roughly six orders of magnitude, from dwarf galaxies to ULIRGS.  The slope 
for the $M_\mathrm{brightest}$ vs.\ log SFR relationship is $-1.73 \pm 0.08$, 
similar to both the $M_\mathrm{brightest}$ vs.\ log~N value 
($-1.97 \pm 0.22$) and the prediction from Larsen 2009 ($-1.82$) discussed above.

The scatter for the LIRG samples is 0.90 magnitude (with IC5283 and
Arp~220 removed), considerably larger than for our spiral sample (i.e., 0.41 
magnitude in Figure~7).  This is partly due to the higher levels of extinction 
present in many of the LIRG galaxies, as shown by the corrections for IC5283 
and Arp~220 (Vavilkin, 2011), which bring the galaxies back into line (see 
Figure~10). Another potential concern is the effect of spatial resolution for the 
more distant galaxies. Are some of the more luminous ``clusters'' in distant 
galaxies really several clusters blended together?  Randriamanakoto et~al.\  
(2013) have performed simulations based on degraded images of the Antennae
galaxies which indicates that this is not likely to be a serious problem for the 
modest redshifts of most of our target galaxies.

\section{CONCLUSIONS}

Hubble observations using the ACS/WFC  camera have been used to construct
star cluster luminosity functions for 20 nearby, star-forming galaxies.  
Automatically generated source lists from the Hubble Legacy Archive (HLA) 
were employed for the project.  These catalogs provide the largest set of uniform, 
automatically-generated cluster candidates we are aware of  in the literature at 
present. The primary results are listed below.

1. Comparisons with other recently generated cluster catalogs (e.g., Bastian et~al.\ 
2012; Chandar et~al.\  2014) demonstrate that the HLA-generated catalogs are of 
similar quality, but in general do not go as deep as manually-generated catalogs. 

2. A single power-law of the form $dN/dL \propto L^{\alpha}$ has been used to 
approximate the LF using three different fitting techniques: constant number and 
constant magnitude binning (e.g., see Maiz-Appellaniz \& Ubeda 2005 and Chandar 
et~al.\ 2010 for discussions), and a maximum likelihood method that does not require 
binning.  The methods give comparable results, although there may be a tendency for 
the maximum likelihood method to give more negative values of $\alpha$ for the 
steeper LFs.

3. Using the mean from the two methods and the high S/N sample, the average value 
for $\alpha$ is  $-2.37$, with a RMS scatter~= 0.18 when using the F814W (``$I$'') band.  
Our values of $\alpha$ are generally steeper than most past studies, with a difference of
$\delta = 0.12 \pm 0.16$ when comparing galaxies one-to-one.

4. A weak correlation is found for galaxies with high values of the SFR (or 
equivalently galaxies with the brightest clusters or the largest number of clusters) to 
have shallower values of $\alpha$.  The same trend is found for $\alpha$ from composite
``supergalaxies" with different SFRs, strengthening the case for the reality of this
correlation. In addition, the Antennae galaxy (NGC~4038/39), a merging system with a
relatively high star formation rate, has the second flattest luminosity function in the sample.
 
5. A weak correlation may be present between $\alpha$ and Hubble Type in the sense 
that later type galaxies (Sd--Sm) have lower values of $\alpha$. 
However, the cumulative distribution functions (CDFs) show
mixed results, hence this result should be considered tentative.

6. While there appear to be some weak correlations, the relative similarity in the values 
of $\alpha$ for a  large number of star-forming galaxies suggests that, to first order, the 
LFs are fairly universal. This is similar to results for mass functions (e.g., Whitmore 
et~al.\ 2010 and Fall and Chandar 2012---but see also Bastian et~al.\ 2012 and Chandar 
et~al.\ 2014).

7. An exercise using larger aperture photometry (radii~= 7~pixels) shows that 
the use  of mean aperture corrections for small aperture photometry does not affect our 
results in a substantial way. 

8. Based on both the LFs and the cumulative distribution function (CDFs) of composite
``super-galaxies,'' we find some evidence for a downturn at the bright end of the 
luminosity functions, although it only pertains to about 1\% of the clusters.

9. The $M_\mathrm{brightest}$ vs.\ log~N relation shows a small RMS scatter (0.4 
to 0.5 mag). It appears that the reason that galaxies with more clusters have brighter 
clusters is primarily a statistical ``size-of-sample'' effect rather than being due to 
differences in the environments of starburst and quiescent galaxies. This is consistent 
with results found by Whitmore (2003) and Larsen (2002).  The results for the  
$M_\mathrm{brightest}$ vs.\ log SFR relationship are similar, with even a smaller 
scatter ($\approx$ 0.4 mag). 

10. The sample has been increased by including  observations of both dwarf galaxies 
and LIRGS from studies by Bastian et~al.\ (2008) and Vavilkin (2010). This shows 
that the $M_\mathrm{brightest}$ vs.\ log SFR correlation extends over roughly six 
orders of magnitudes. However, higher levels of extinction appear to  lead to larger 
scatter (0.9 magnitude) for the LIRG  sample.

The photometric catalogs used in this study are available at:\hfill\break
http://archive.stsci.edu/prepds/starclust-sfgal/.

\acknowledgements
We thank the referee for several useful comments. BCW acknowledges the use of STScI grant GO-11781 and the summer student program at STScI for support of this project. 

\clearpage

\renewcommand{\thefigure}{\arabic{figure}\alph{subfigure}}
\setcounter{subfigure}{1}

\setcounter{subfigure} {1}
\begin{figure}
\epsscale{.725}
\plotone{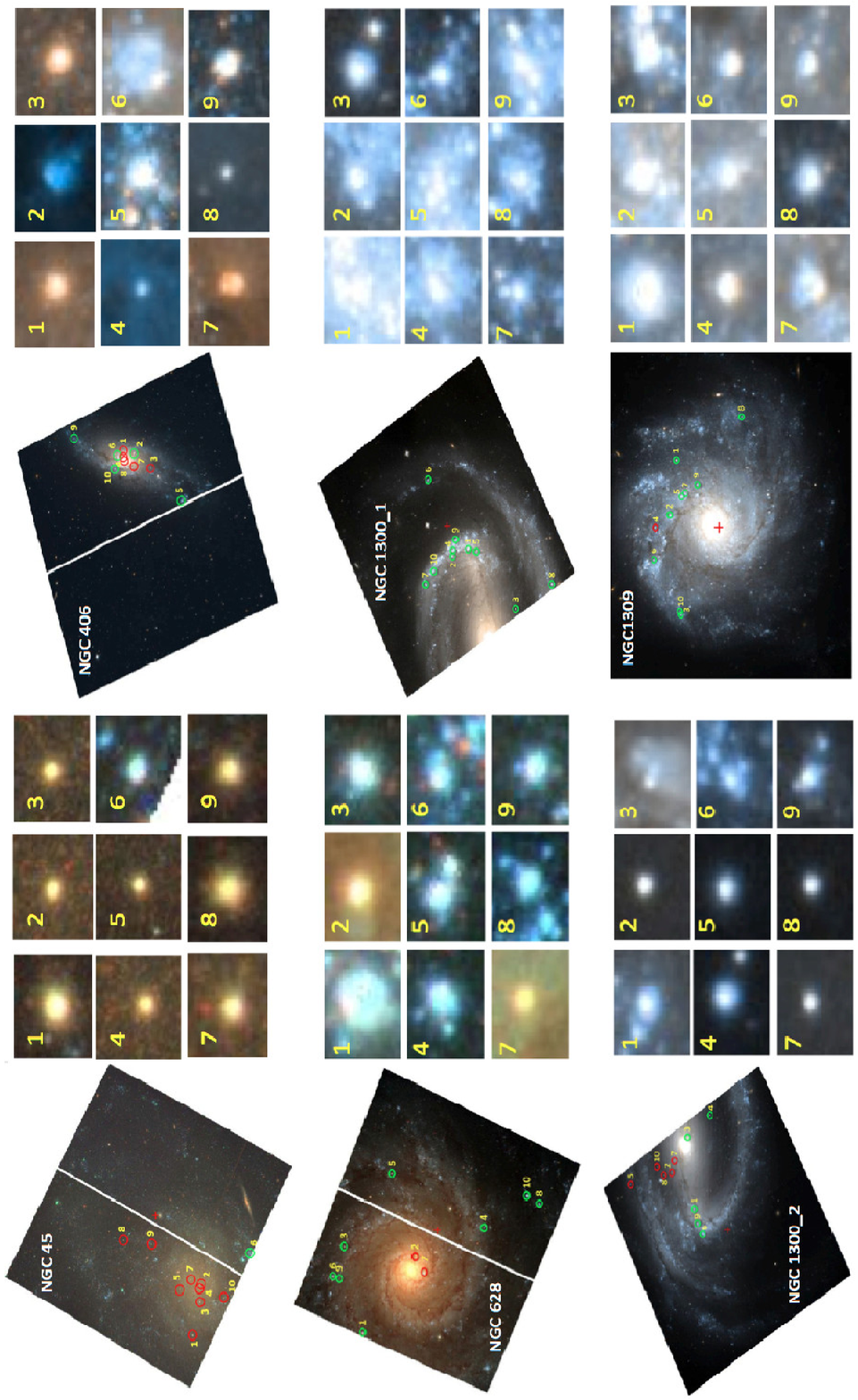}
\caption{Color images of the 20 galaxies used in this work.  Red circles show young (blue) clusters while green circles show old clusters. $5\times5\arcsec$ cutouts of the 9~brightest sources in each HLA catalog are also shown.}  
\end{figure}

\addtocounter{figure}{-1}
\addtocounter{subfigure}{1}
\begin{figure}
\epsscale{.9}
\plotone{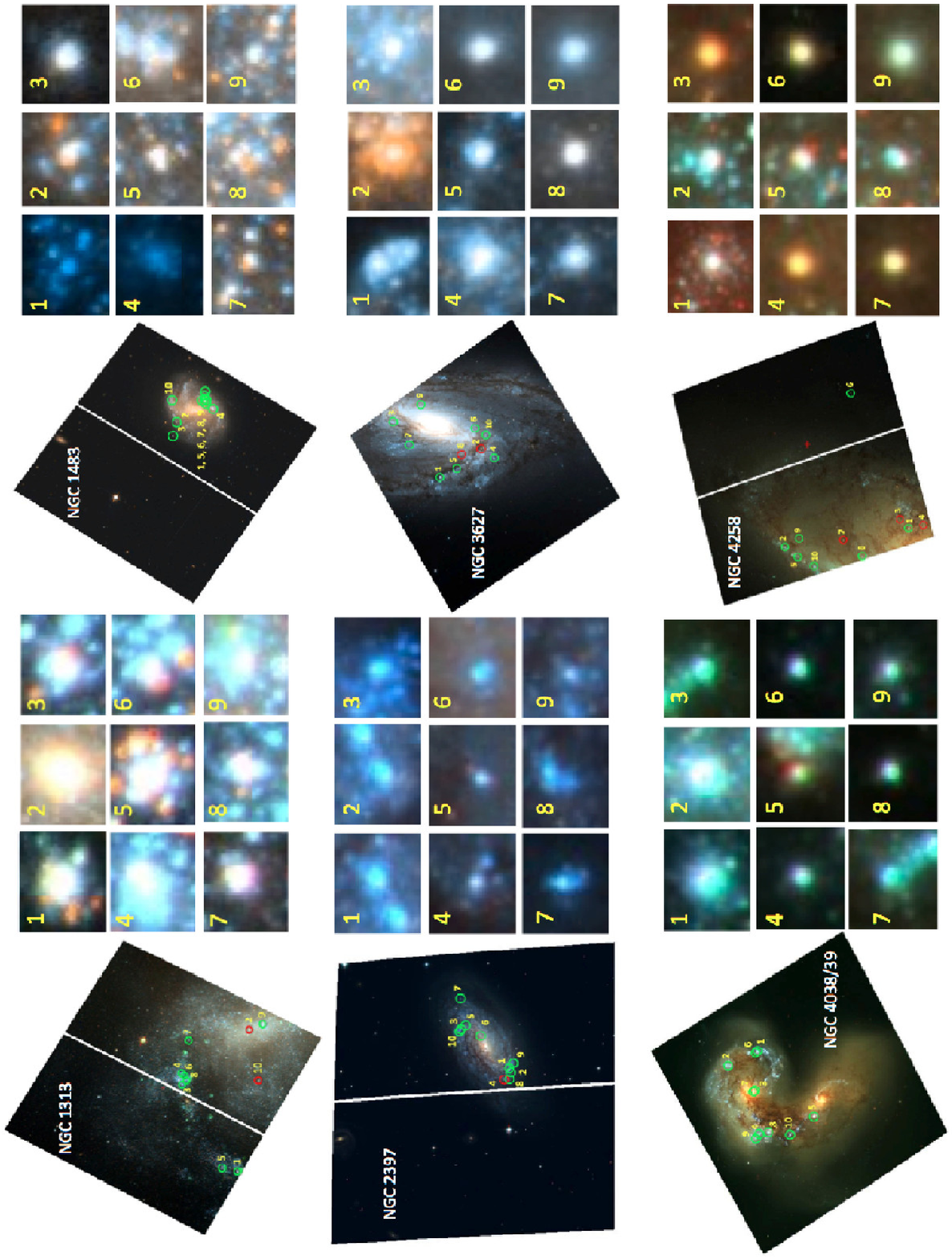}
\caption{Continued.}
\end{figure}

\addtocounter{figure}{-1}
\addtocounter{subfigure}{1}
\begin{figure}
\plotone{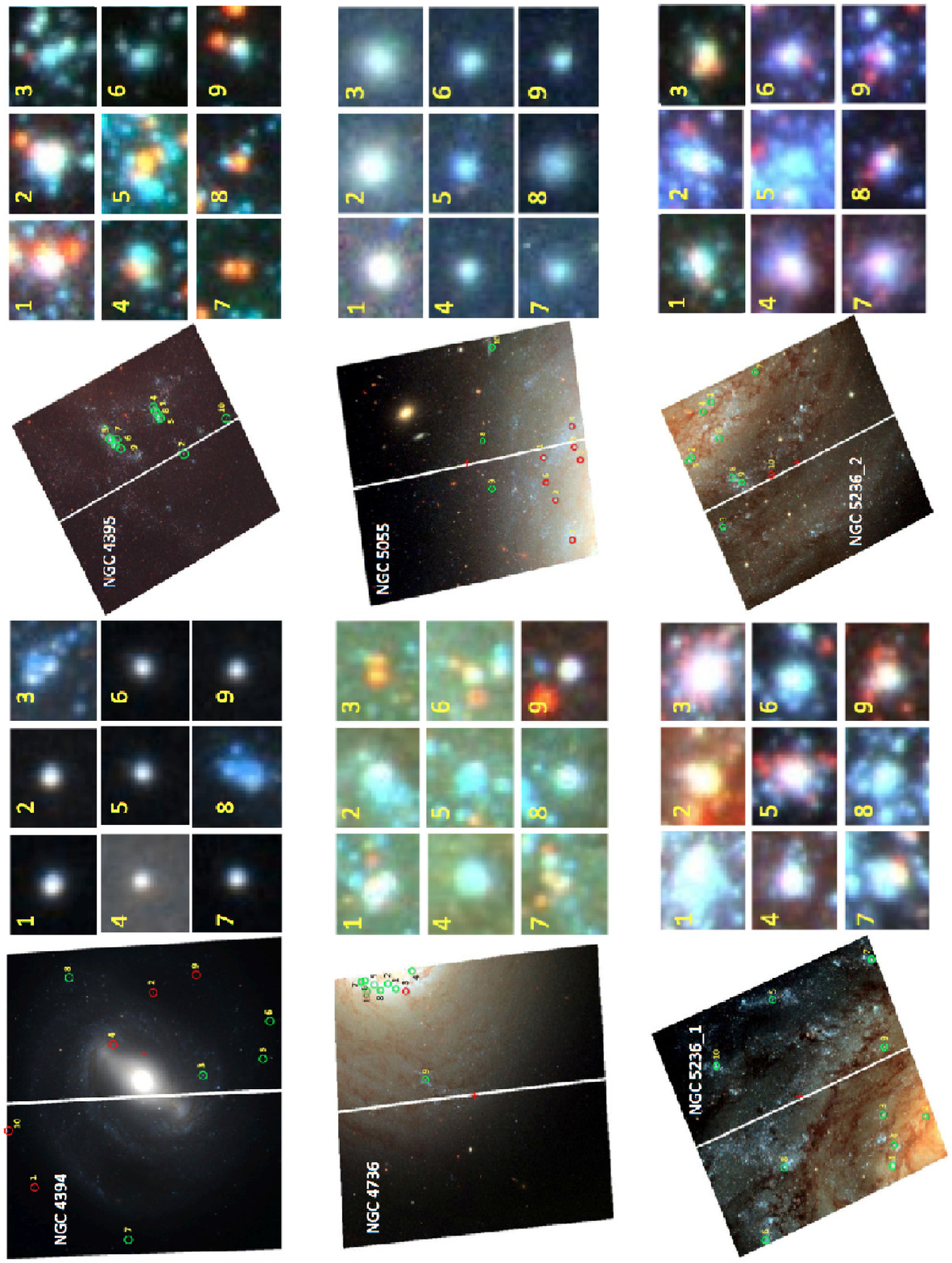}
\caption{Continued.}
\end{figure}

\addtocounter{figure}{-1}
\addtocounter{subfigure}{1}
\begin{figure}
\plotone{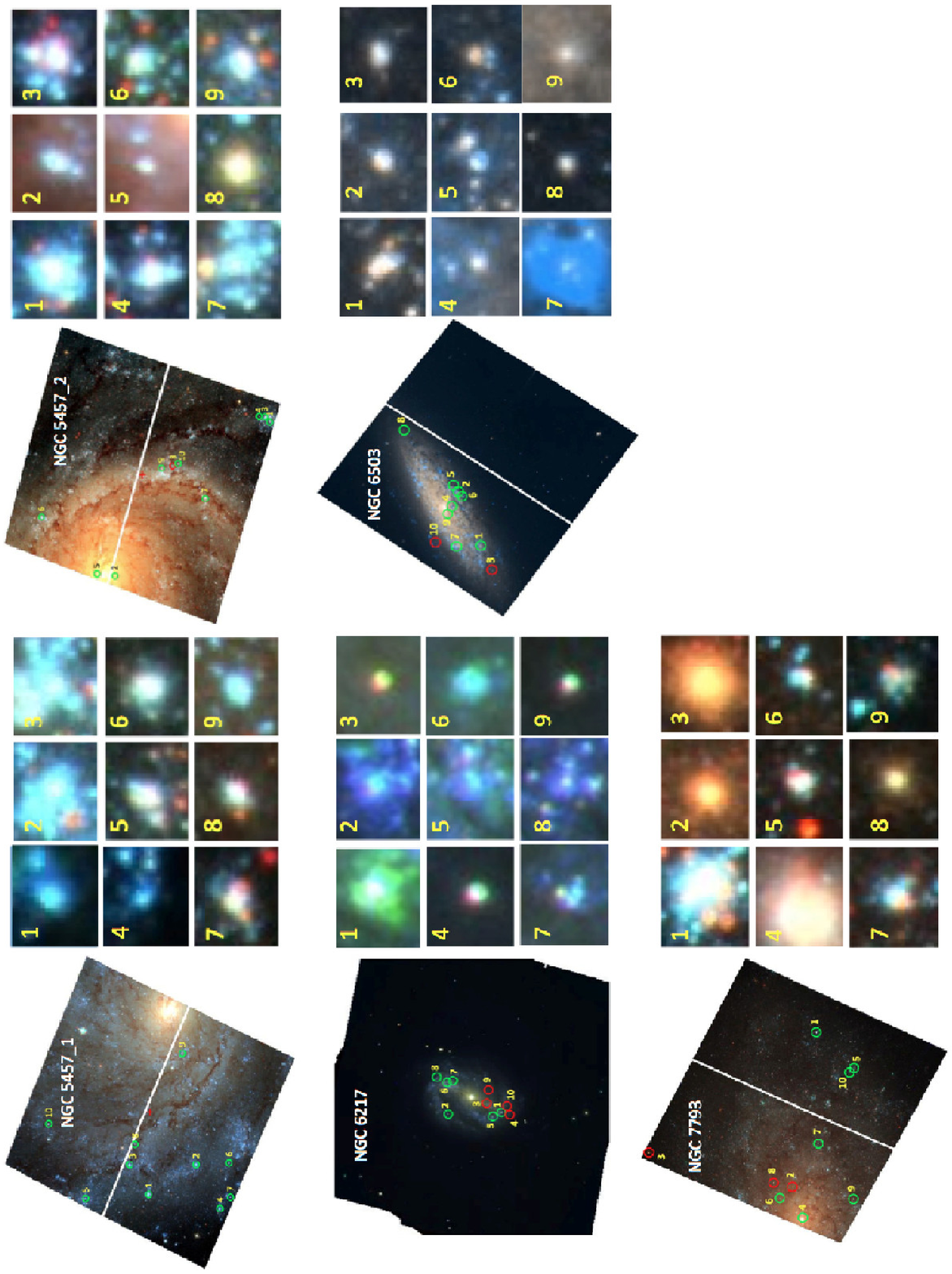}
\caption{Continued.}
\end{figure}
\clearpage
\renewcommand{\thefigure}{\arabic{figure}}

\begin{figure}
\epsscale{.525}
\plotone{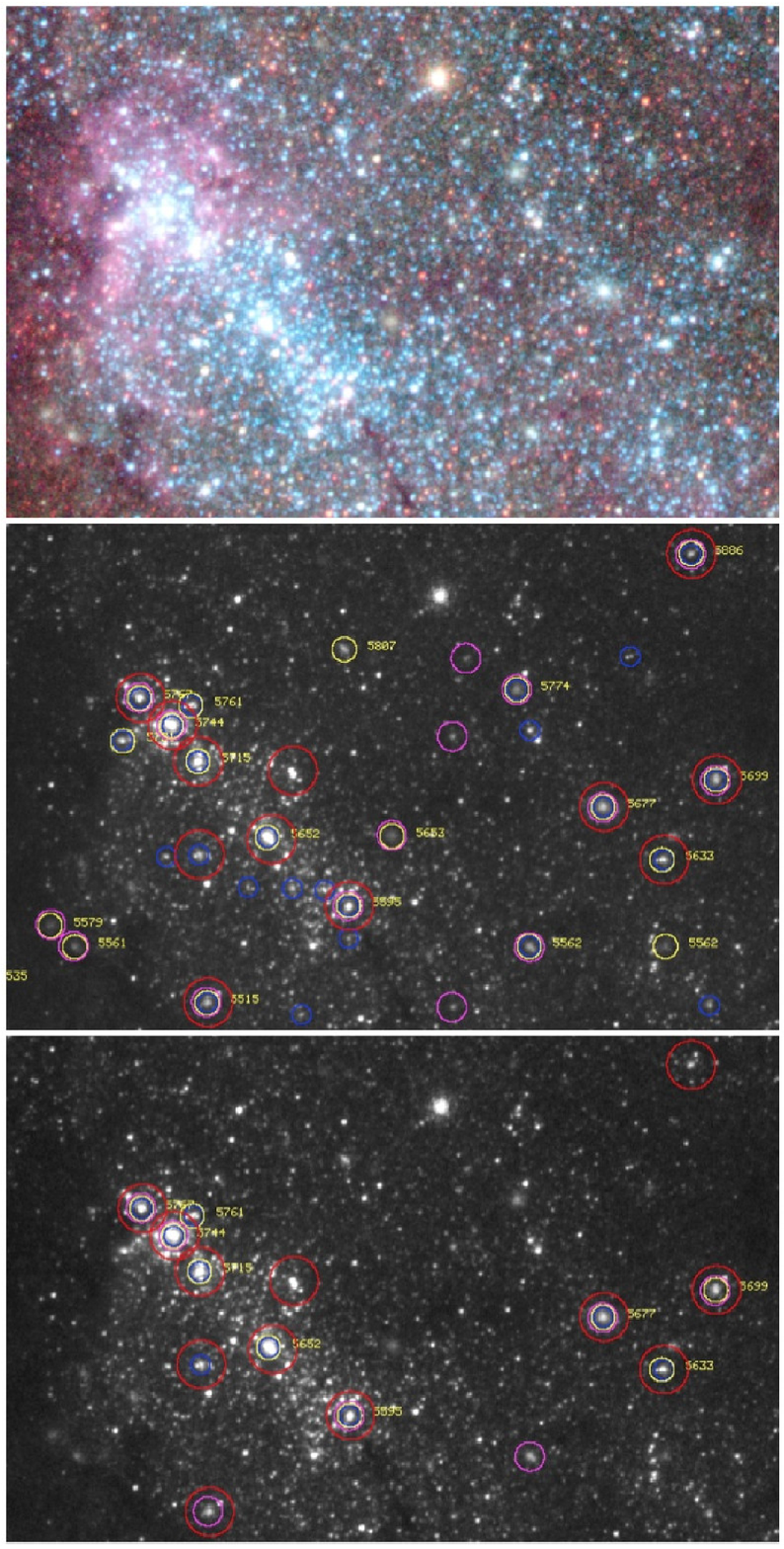}
\caption {Comparison between catalogs discussed in Section~3. The top panel shows a color image of the region in M83 selected to make the comparison.  The middle panel shows a comparison of the HLA-generated cluster catalog used in the present study (large red circles) with the other three catalogs (Chandar-automatic in blue, Chandar-manual in yellow, Bastian-hybrid in purple---see Chandar et~al.\ 2013 for a similar figures and detailed
description). The bottom panel imposes similar magnitude thresholds in order to facilitate a more equitable comparison.}
\end{figure}

\begin{figure}
\epsscale{1.0}
\plotone{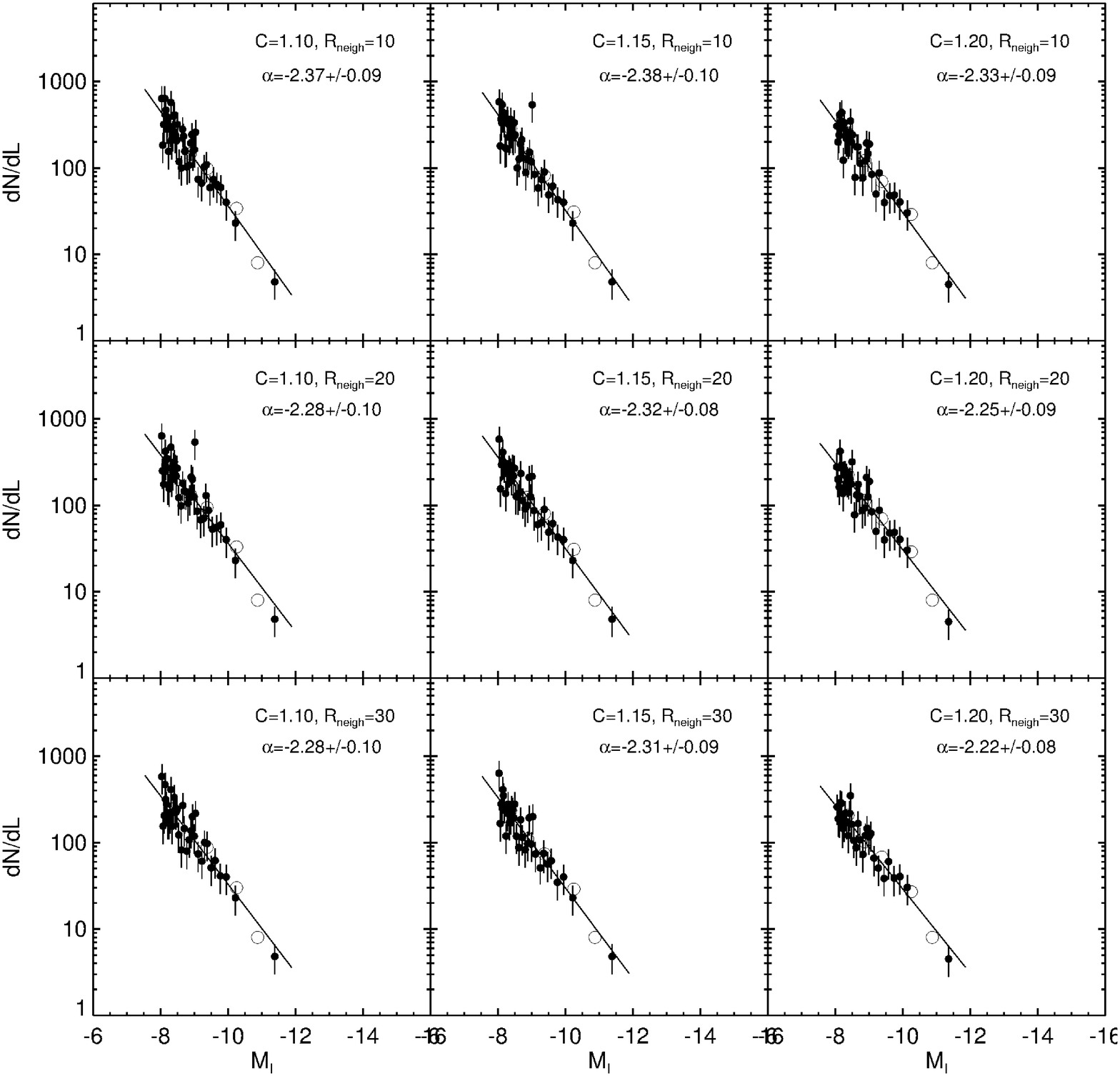}
\caption{LFs for $3\times3$ grid of $C$ and UNIQPOS parameters for M83. Open circles are for constant magnitude binning while filled circles are for constant number binning.  See Section~3 for details.}
\label{fig:lf}
\end{figure}

\begin{figure}
\epsscale{.7}
\plotone{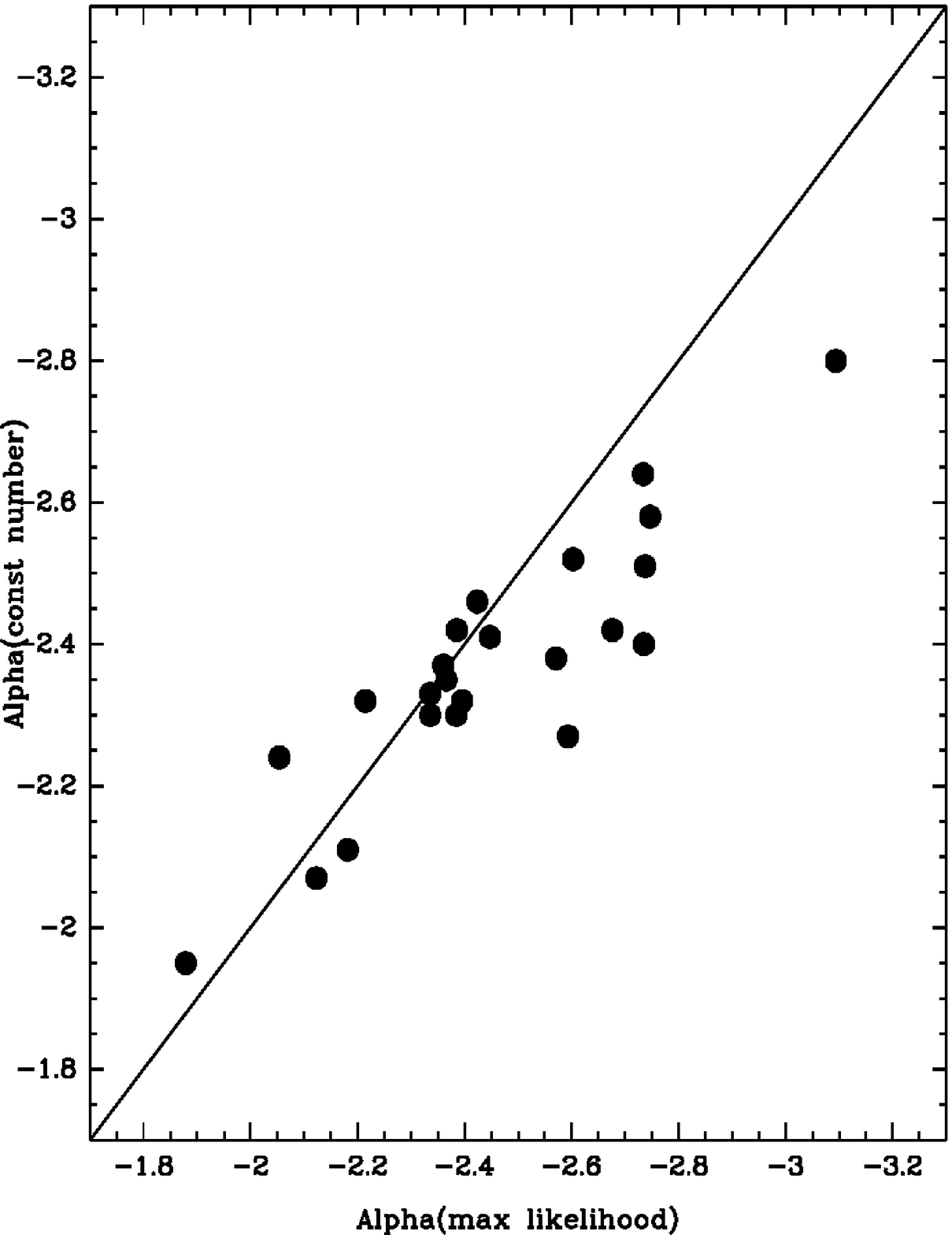}
\caption{Comparison between values of $\alpha$ determined using the maximum likelihood and the constant number binning methods.}
\label{fig:cp}
\end{figure}

\begin{figure}
\epsscale{1.0}
\plotone{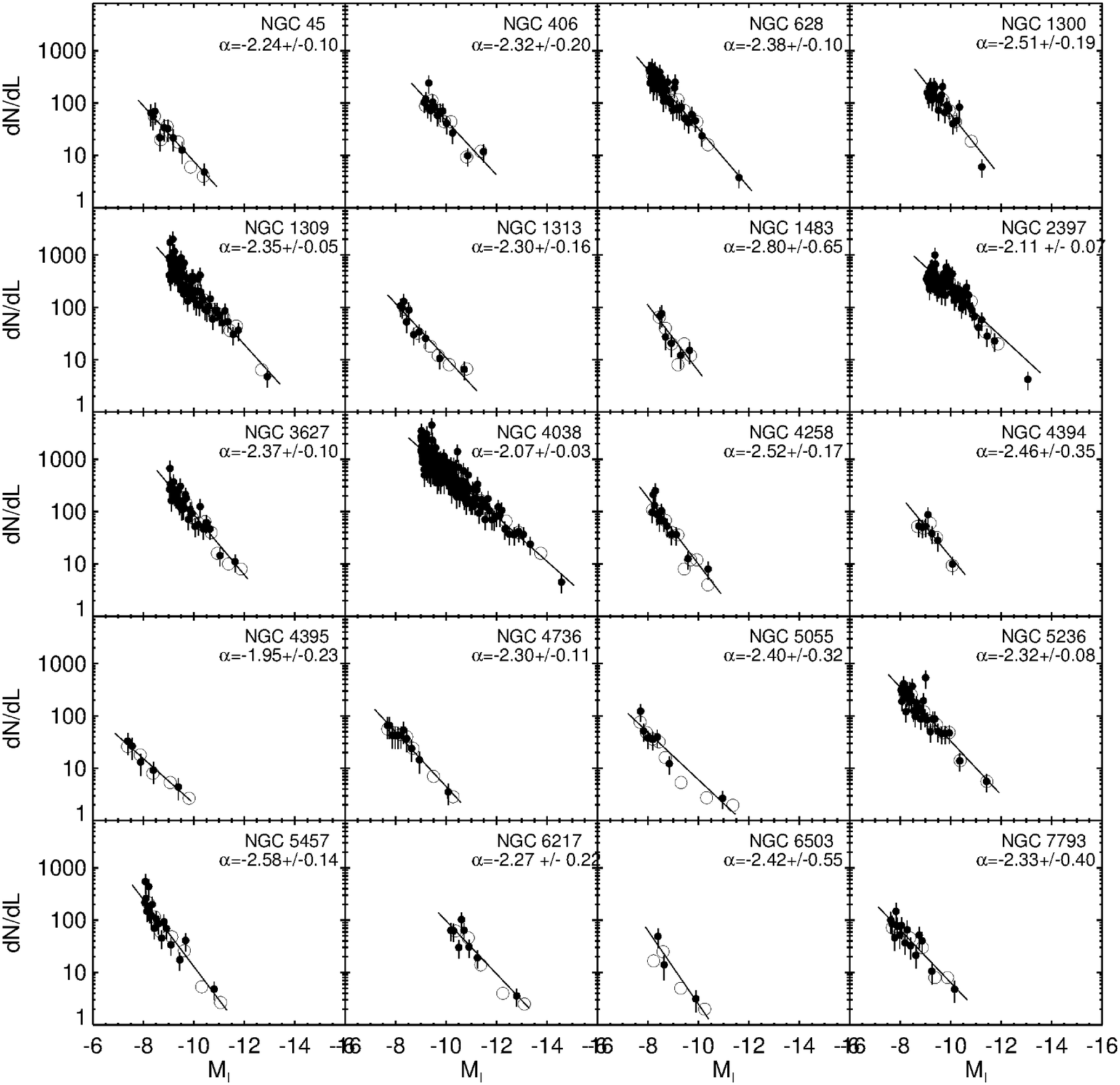}
\caption{LFs for all 20 galaxies in our sample. Filled circles show the results using constant number per bin while open symbols show the results using constant magnitude bins. The values for $\alpha$ are for the constant number binning.}
\label{fig:lf}
\end{figure}

\begin{figure}
\epsscale{.85}
\plotone{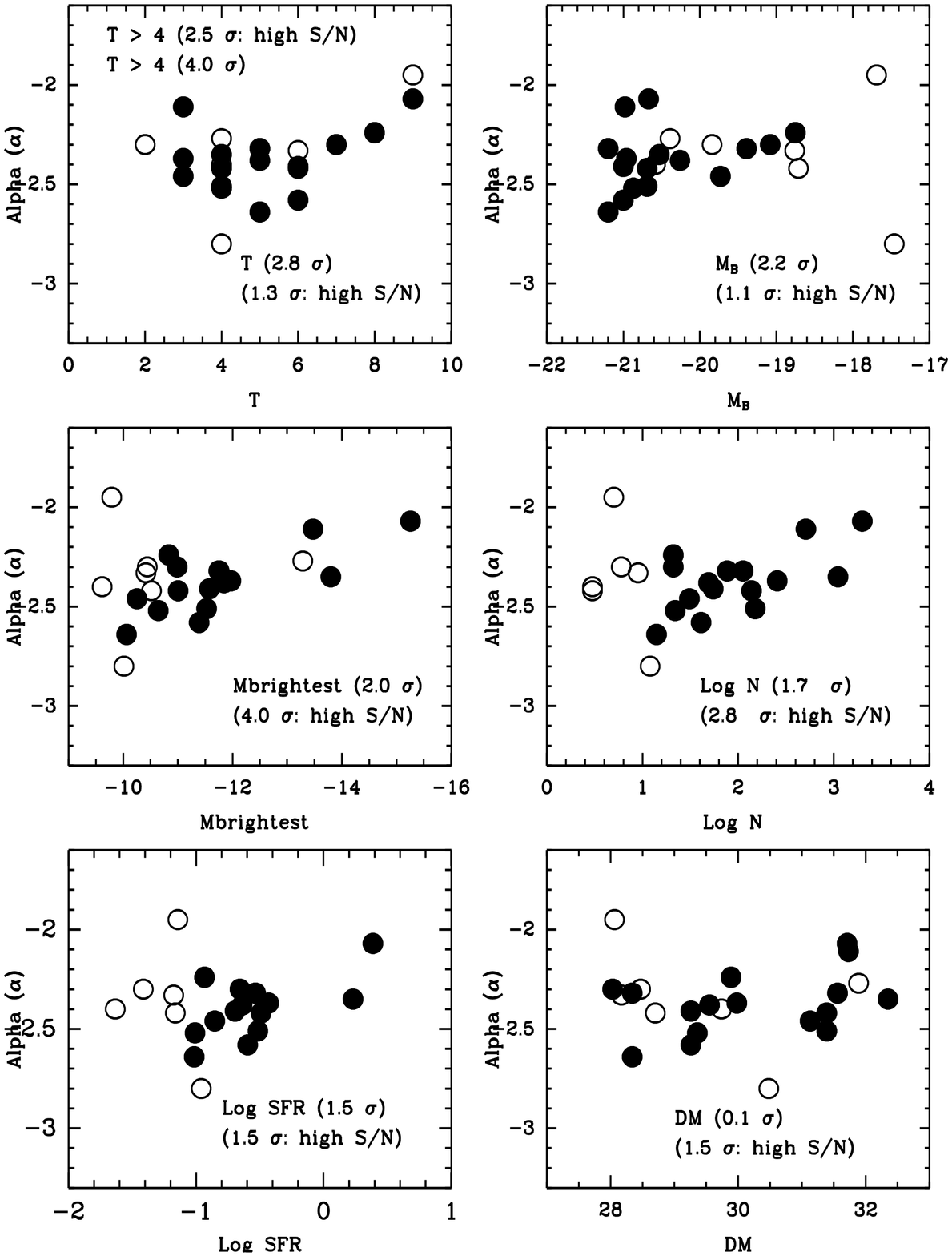}
\caption{Correlation search using the $\alpha$ determinations from the constant number binning. Filled circles show the high S/N data (i.e., log N $> 1.20$ for $M_I -9$ mag limit)  while open circles show the low S/N data. See Section~5.1 for details.}
\label{fig:cs2}
\end{figure}

\begin{figure}
\epsscale{.70}
\plotone{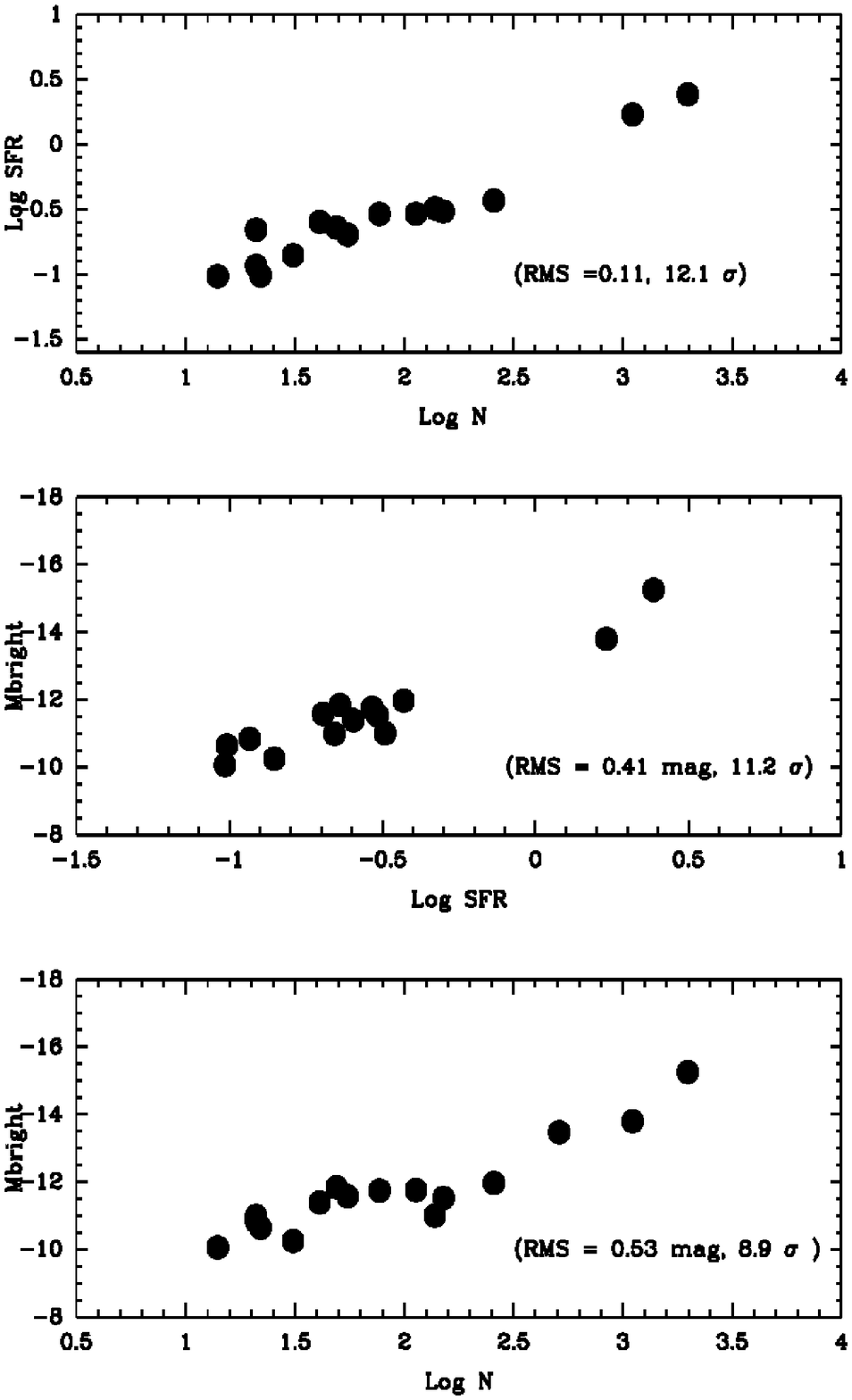}
\caption{The correlations between $M_\mathrm{brightest}$, log(N), and log(SFR).}
\label{fig:cs5}
\end{figure}

\begin{figure}
\epsscale{.85}
\plotone{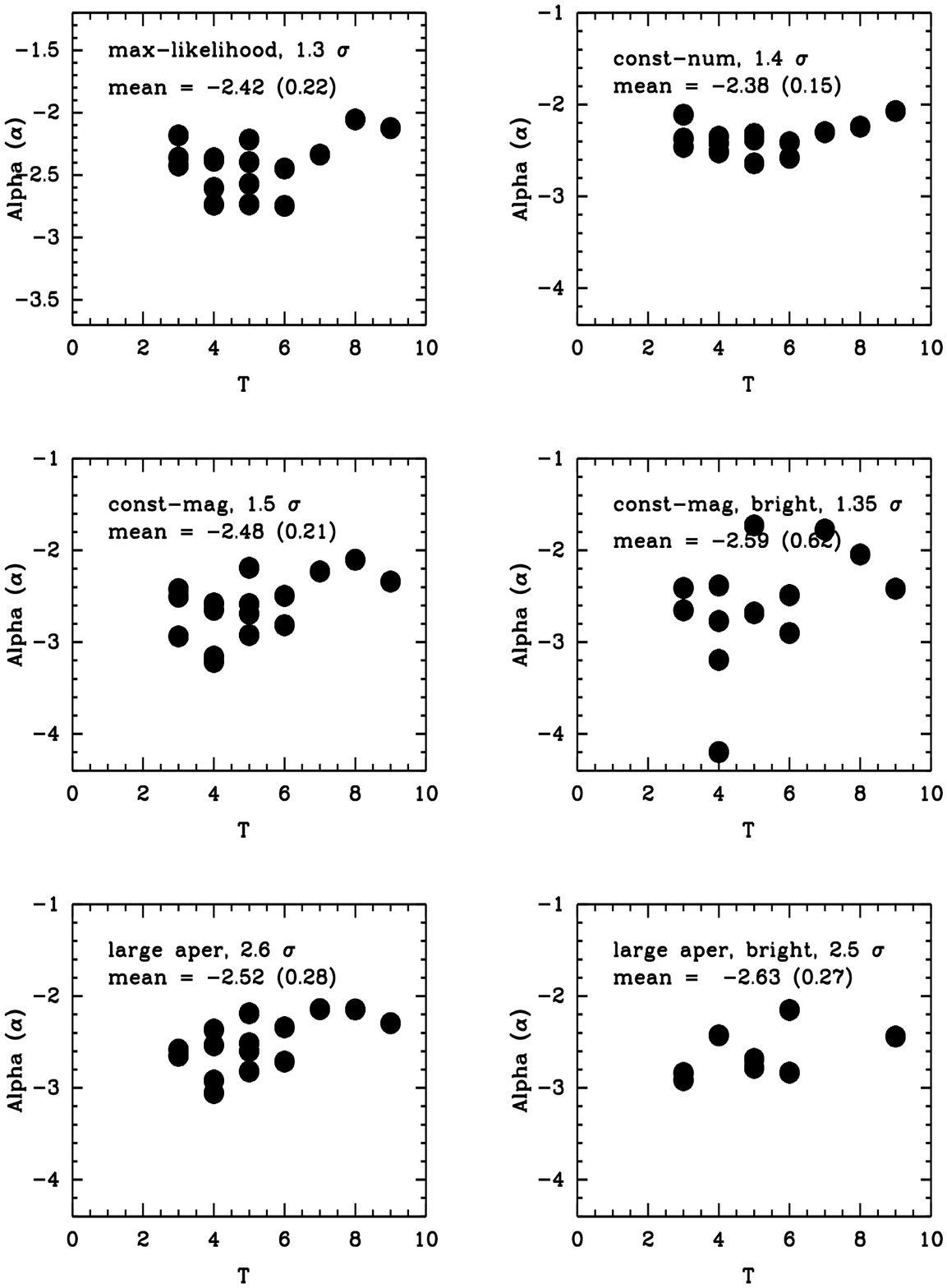}
\caption{The correlations between Alpha ($\alpha$)  and  Hubble type (T) using three different fitting methods (i.e., maximum-likelihood, constant number and constant magnitude); 7~pixels photometry (i.e.,  ``large aper''); and the brighter part of the luminosity function (i.e., ``bright''). See Section~3 for details. }
\label{fig:cs5}
\end{figure}

\begin{figure}[ht!]
\epsscale{.4}
\plotone{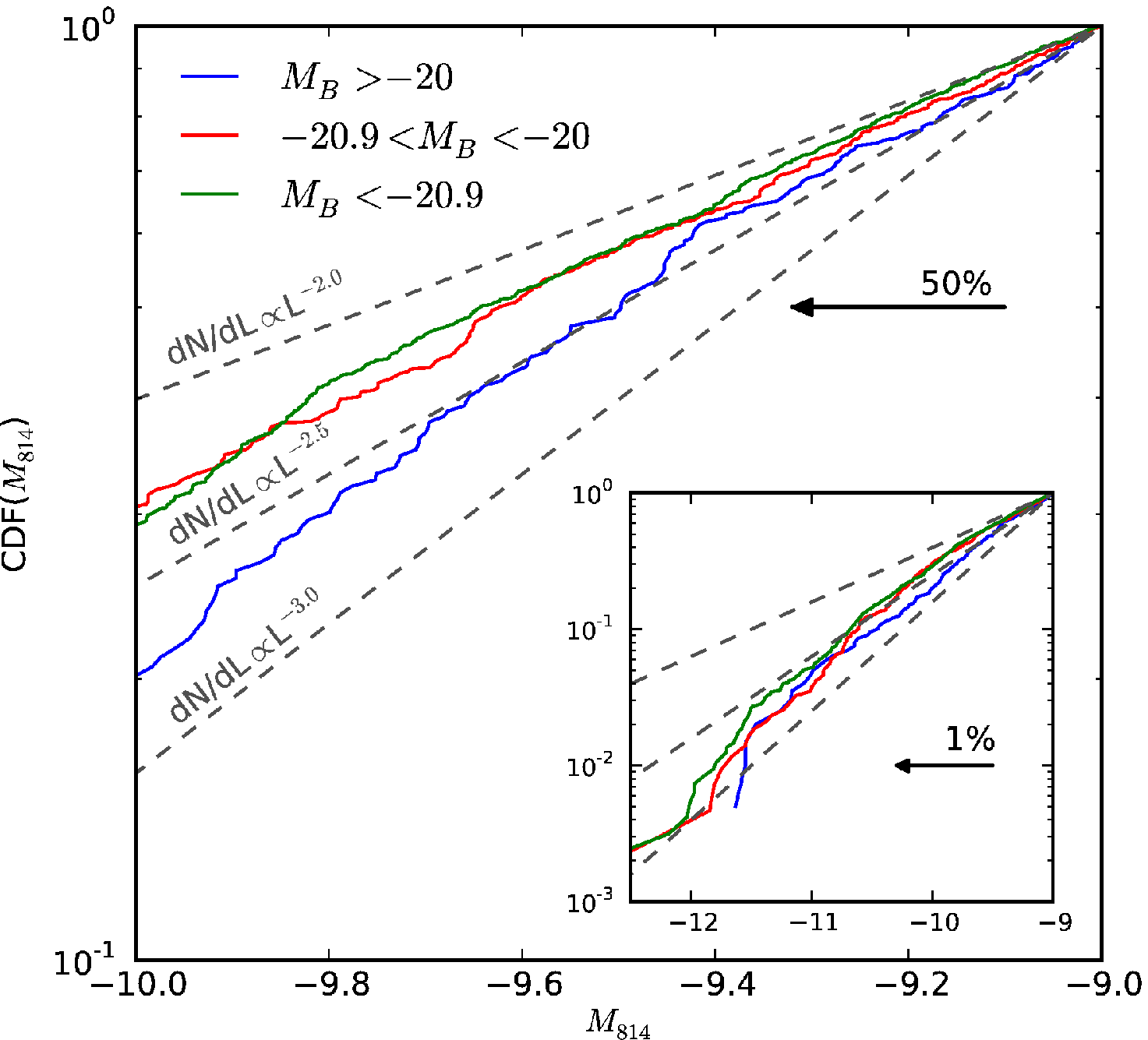}\\
\smallskip
\plotone{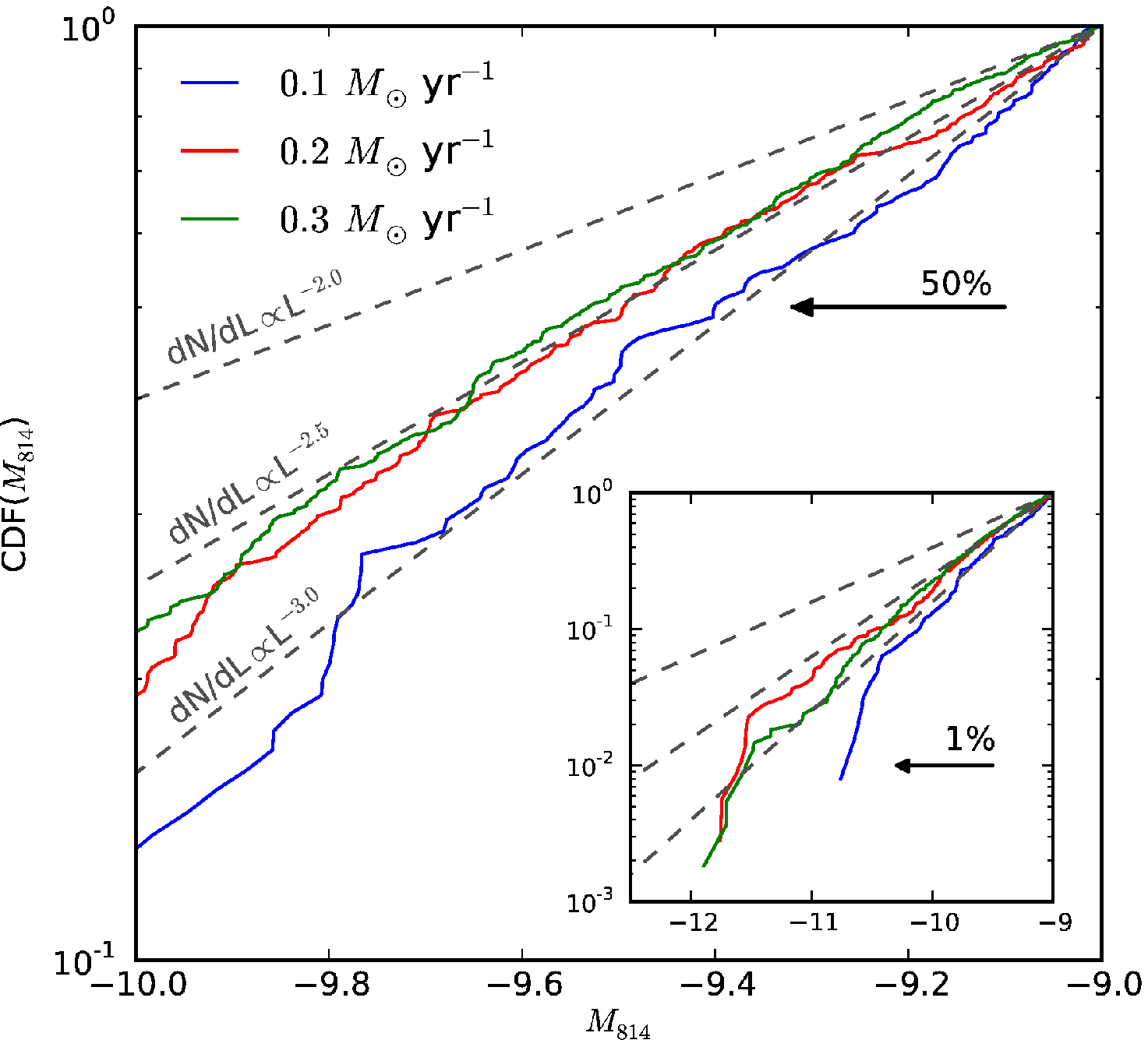}\\
\smallskip
\plotone{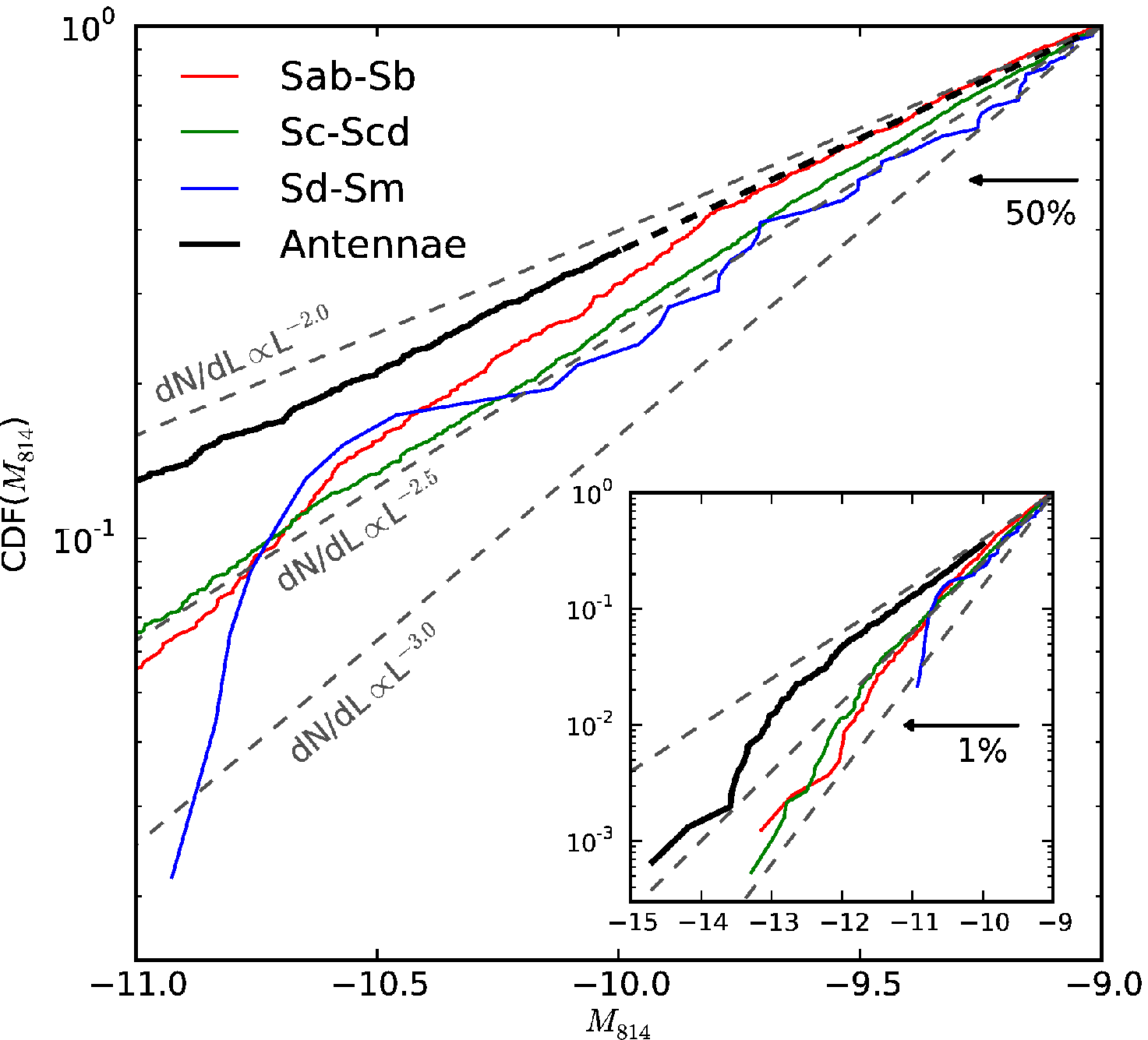}
\caption{Each plot shows the Cumulative Distribution Functions (CDF) for the composite 
``supergalaxies" grouped by: top -- absolute B magnitude, middle -- SFR,  bottom -- Hubble type.  Lines with constant values of  $\alpha = -2$, $-2.5$ and $-3$ are included 
for comparison. See Section~5.2 for details.} 
\end{figure}

\begin{figure}
\centering
\includegraphics[width=3.in, angle= 90]{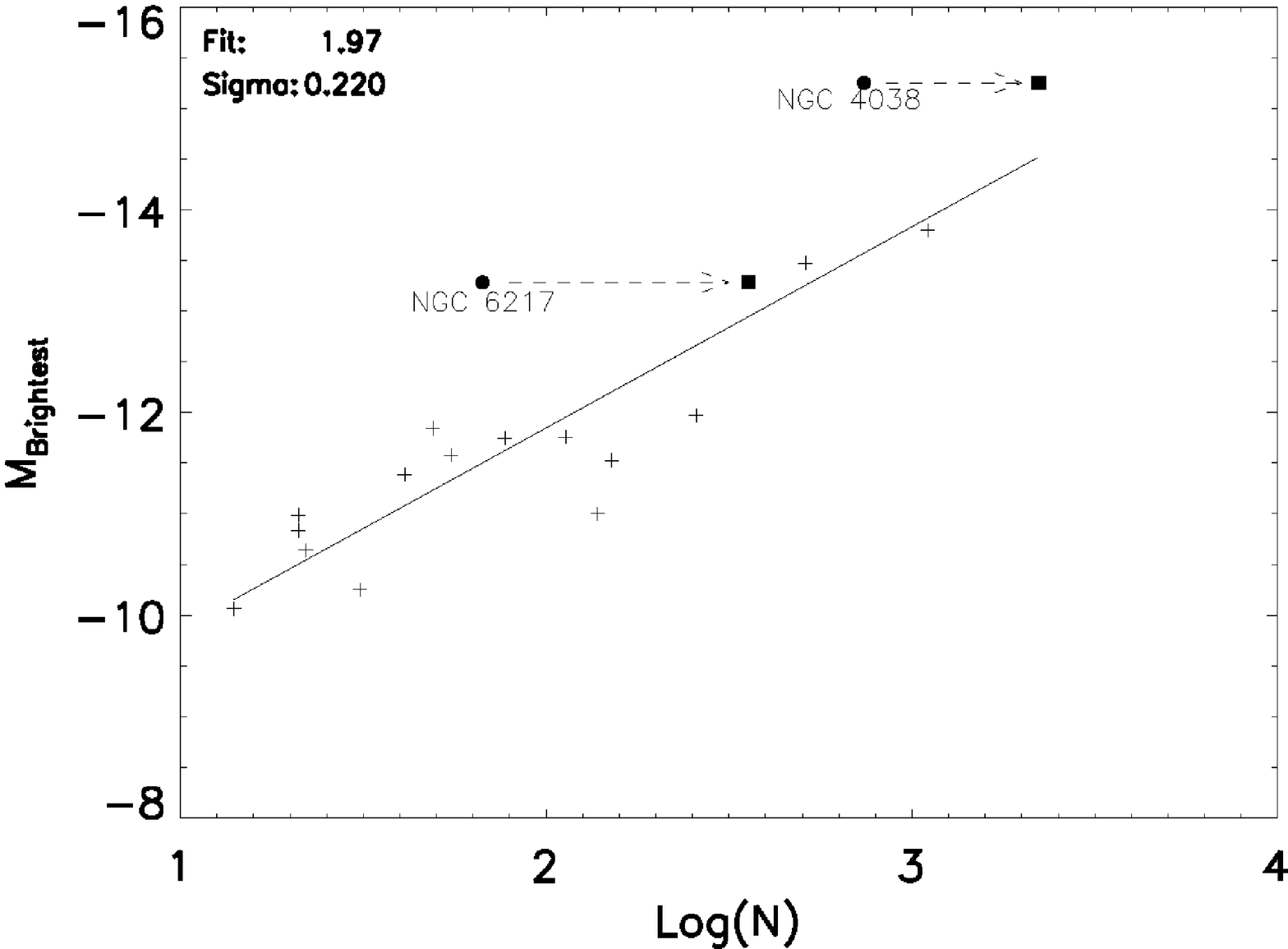}\\
\includegraphics[width=3.in, angle= 90]{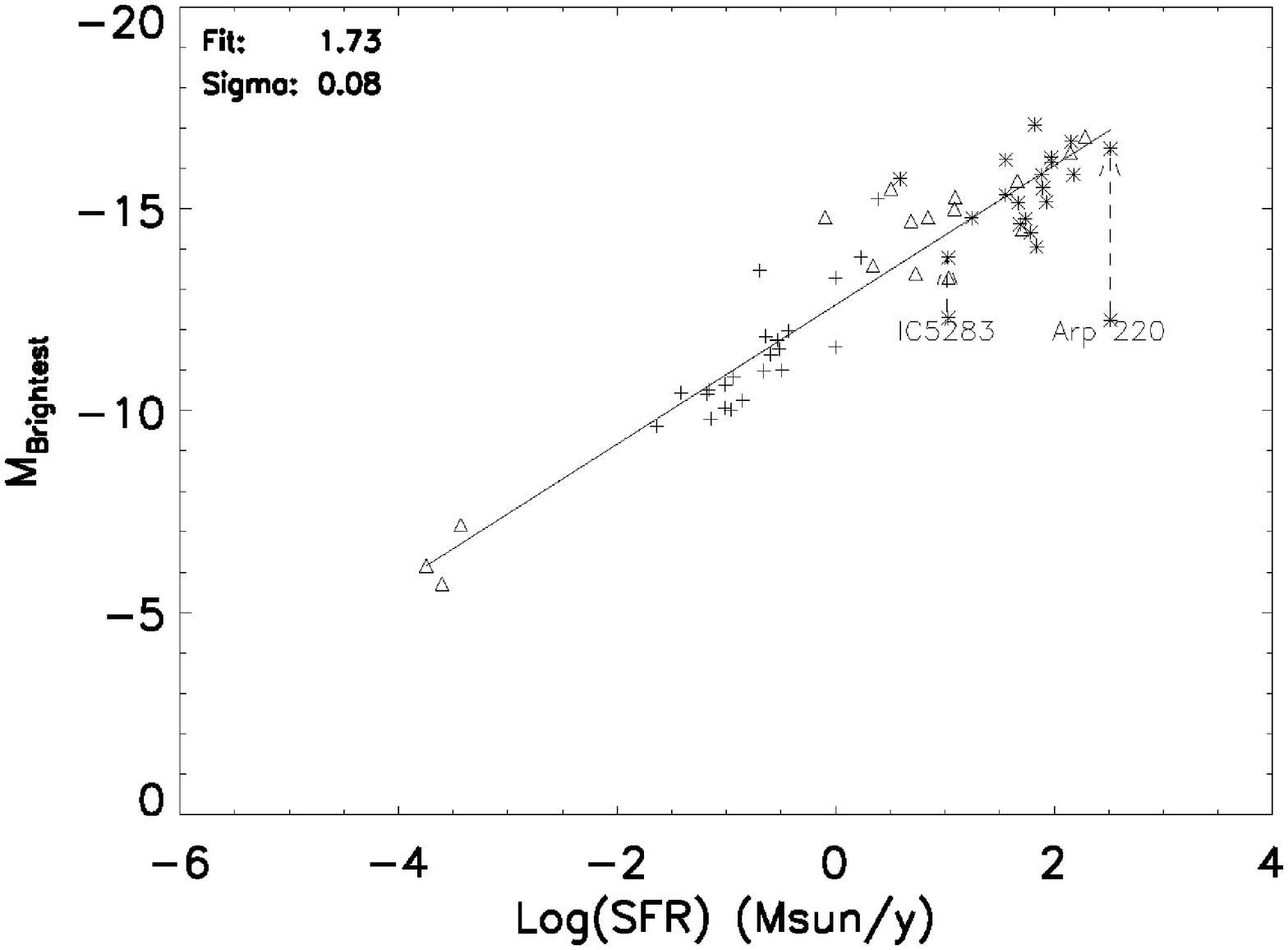}
\caption{Log SFR vs.\ brightest cluster for the sample galaxies. The galaxies
from the Vavilkin (2011) sample are marked with asterisks; the triangles are data from
Bastian (2008). Data from the current sample are marked with a plus sign. See Section~5.3 for details. }
\end{figure}
\clearpage



\begin{deluxetable}{llcccclc}
\tabletypesize{\footnotesize}
\rotate
\tablewidth{0pt}
\tablecaption{Basic Properties of Target Galaxies\label{tab:obs}}
\tablehead{
\colhead{Galaxy} & \colhead{Type} & \colhead{$m-M$} & \colhead{$A_{I}$} & \colhead{$M_B$} & \colhead{Proposal ID visit}  & \colhead{Filters} & \colhead{SFR}
}
\startdata
NGC 45 &SA(s)dm&29.89&0.040&$-$18.75 &9774 03 &\emph{F336W}, F435W, F555W, F814W  &0.1165 \\
NGC 406 &SA(s)c&31.56&0.046&$-$19.39 &9395 03 &F435W, F555W, F814W  &0.2931 \\
NGC 628 &SA(s)c&29.55&0.135&$-$20.26 &\llap{1}0402 23 &\emph{F336W}, F435W, F555W, F658N, F814W  &0.2295 \\
NGC 1300 (field 1) &SB(rs)bc&31.39&0.059&$-$20.69 &\llap{1}0342 02 &F658N, F814W  &0.3051 \\
NGC 1300 (field 2)&SB(rs)bc&31.39&0.059&$-$20.69&\llap{1}0342 04 &F658N, F814W  &0.3225 \\
NGC 1309 &SA(s)bc&32.35&0.077&$-$20.53 &\llap{1}0497 10 &F555W, F814W  &1.7038 \\ 
NGC 1313 &SB(s)d &28.03&0.212&$-$19.08 & 9774 05 &\emph{F336W}, F435W, F555W, F814W  &0.2205 \\ 
NGC 1483 &SB(s)bc&30.48&0.014&$-$17.46 &9395 05 &\emph{F336W}, F435W, F555W, F658N, F814W  &0.1098 \\ 
NGC 2397&SB(s)b&31.73&0.398&$-$20.98 &\llap{1}0498 02&F435W, F555W, F814W  & \nodata \\
NGC 3627 &SAB(s)b&29.98&0.063&$-$20.96 &\llap{1}1575 01 &\emph{F336W}, F435W, F555W, F658N, F814W  & 0.3709 \\ 
NGC 4038 & SB(s)mpec&31.71&0.090&$-$20.67 &\llap{1}0188 10 &\emph{F336W}, F435W, F550M, F814W  &2.4316 \\ 
NGC 4258 & SAB(s)bc&29.36&0.031&$-$20.87 &9810 12 &F435W, F555W, F814W  &0.0981 \\ 
NGC 4394 & (R)SB(r)b &31.13 &0.059&$-$19.73 &\llap{1}0515 17 & F475W, F814W  &0.1402 \\ 
NGC 4395& SA(s)m&28.06&0.033&$-$17.69 &9774 0b &\emph{F336W}, F435W, F555W, F814W  &0.0719 \\ 
NGC 4736&(R)SA(r)ab&28.47&0.034&$-$19.84 &\llap{1}0402 07 &\emph{F336W}, F435W, F555W, F658N, F814W  &0.0385 \\ 
NGC 5055&SA(rs)bc&29.74&0.034&$-$20.58 &\llap{1}0402 18 &\emph{F336W}, F435W, F555W, F606W, F814W  &0.0232 \\ 
NGC 5236 (field 1)&SAB(s)c&28.34&0.128&$-$21.20 &9774 0f &\emph{F336W}, F435W, F555W, F814W  &0.2915 \\
NGC 5236 (field 2)&SAB(s)c&28.34&0.128&$-$21.20 &9774 0h &\emph{F336W}, F435W, F555W, F814W  &0.0967 \\ 
NGC 5457 (field 1)&SAB(rs)cd&29.26&0.017&$-$21.00 &9490 01 &\emph{F336W}, F435W, F555W, F814W  &0.2541 \\ 
NGC 5457 (field 2)&SAB(rs)cd&29.26&0.017&$-$21.00 &9490 02 &\emph{F336W}, F435W, F555W, F814W  &0.2021 \\
NGC 6217&(R)SB(rs)bc&31.89&0.085&$-$20.39 &\llap{1}1371 03&F658N, F814W  &\nodata \\
NGC 6503 &SA(s)cd&28.70&0.062&$-$18.71 &9293 11 &F658N, F814W  &0.0688 \\ 
NGC 7793& SA(s)cd&28.17&0.038&$-$18.76 &9774 0l &\emph{F336W}, F435W, F555W, F814W  &0.0667 \\
\enddata
\tablecomments{All observations are from the ACS except F336W (in italics) which is from the WFPC2. F336W is included as reference information but not used in the paper.}
\end{deluxetable}


\begin{deluxetable}{lcccccc}
\tabletypesize{\footnotesize}
\rotate
\tablewidth{0pt}
\tablecaption{Summary of Cluster Catalogs\label{tab:gals}}
\tablehead{
\colhead{Galaxy}  & \colhead{Brightness} &
\colhead{Concentration} & \colhead{Uniqpos}  &
\colhead{$M_\mathrm{brightest}$} & \colhead{log N @ Brightness}  & \colhead{log N @ $M_I = -9$}\\
\colhead{} & \colhead{Limit, $M_I$} &\colhead{Index, $C$} & \colhead{$R_\mathrm{neighbor}$}  & \colhead{} &\colhead{Limit}  & \colhead{Limit} 
}
\startdata 
NGC 45 & \phn$-$8.0 & 1.15& 10 & $-$10.83& 1.79& 1.32\\
NGC 406 & \phn$-$9.0& 1.10& 10 & $-$11.75& 2.05& 2.05\\
NGC 628/M74 & \phn$-$8.0& 1.15& 10 & $-$11.84& 2.29& 1.69\\
NGC 1300 (field 2) & \phn$-$9.0 &1.10& 10 & $-$11.00& 2.13& 2.13\\
NGC 1300 (field 1) & \phn$-$9.0 &1.10& 10 & $-$11.52& 2.17& 2.17\\
NGC 1309 & \phn$-$9.0 & 1.10 & 10 & $-$13.80& 3.04& 3.04\\ 
NGC 1313 & \phn$-$8.0 & 1.25 & 25 & $-$10.98& 1.91& 1.32\\ 
NGC 1483 & \phn$-$8.5 & 1.13 & 15 & $-$10.01& 1.61& 1.07\\ 
NGC 2397 & \phn$-$9.0 & 1.10 &10 & $-$13.47& 2.70& 2.70\\
NGC 3627/M66 & \phn$-$9.0 & 1.10 & 10 & $-$11.97& 2.40& 2.40\\ 
NGC 4038 & $-$10.0 & 1.10 & 10 & $-$15.25& 2.86& \nodata\\ 
NGC 4258 & \phn$-$8.0& 1.15 & 10 & $-$10.64& 2.04& 1.34\\ 
NGC 4394 & \phn$-$8.5& 1.10 & 10 & $-$10.25& 1.74& 1.49\\ 
NGC 4395 & \phn$-$7.0& 1.30 & 30 & \phn$-$9.79&1.43 & 0.69\\ 
NGC 4736 & \phn$-$7.5& 1.14 & 20 & $-$10.44& 1.79& 0.77\\ 
NGC 5055/M63 & \phn$-$7.5 & 1.15 & 10 & \phn$-$9.61& 1.68& 0.47\\ 
NGC 5236 (field 1)/M83 & \phn$-$8.0 & 1.15 & 20 & $-$11.74& 2.42& 1.88\\
NGC 5236 (field 2)/M83 & \phn$-$8.0 & 1.15 & 20 & $-$10.06& 1.88& 1.14\\ 
NGC 5457 (field 1)/M101& \phn$-$8.0 & 1.15& 20 & $-$11.38& 2.29& 1.60\\ 
NGC 5457 (field 2)/M101& \phn$-$8.0 & 1.15& 20 & $-$11.57& 2.30& 1.74\\
NGC 6217 & $-$10.0&1.10 & 10 & $-$13.28& 1.82& \nodata\\
NGC 6503 & \phn$-$8.0& 1.20 & 20 & $-$10.51& 1.39& 0.47\\ 
NGC 7793 & \phn$-$7.5& 1.24 & 30 & \phn$-$9.65& 1.36& 0.60\\
\enddata
\end{deluxetable}


\begin{deluxetable}{lccc}
\tablewidth{0pt}
\tablecaption{Luminosity Function Fits\label{tab:alpha}}
\tablehead{
\colhead{Galaxy} & \colhead{$\alpha_\mathrm{max-likelihood}$} 
& \colhead{$\alpha_\mathrm{const-number}$} & \colhead{Other $\alpha$}
}
\startdata
NGC 45 &$-2.05\pm0.19$&$-2.24\pm0.10$&$-1.94\pm0.28^{\rm a}$ \\
NGC 406 &$-2.21\pm0.15$&$-2.32\pm0.20$&\nodata \\
NGC 628 &$-2.57\pm0.12$&$-2.38\pm0.10$&$-2.16\pm0.26^{\rm b}$ \\
NGC 1300 (field 1) &$-2.74\pm0.16$&$-2.51\pm0.13$&\nodata \\
NGC 1300 (field 2) &$-2.39\pm0.18$&$-2.42\pm0.18$&\nodata\\
NGC 1309 &$-2.37\pm0.04$&$-2.35\pm0.05$&\nodata\\
NGC 1313 (field 1) &$-2.34\pm0.18$&$-2.30\pm0.16$&$-2.10\pm0.12^{\rm b}$,\\
&&&$-2.08\pm 0.10^{\rm c}$\\
NGC 1483 &$-3.09\pm0.46$&$-2.80\pm0.65$&\nodata\\
NGC 2397 &$-2.18\pm0.06$&$-2.11\pm0.07$&\nodata\\
NGC 3627 &$-2.36\pm0.10$&$-2.37\pm0.10$&$-2.50\pm0.07^{\rm d}$\\
NGC 4038 &$-2.12\pm0.04$&$-2.07\pm0.03$&$-2.13\pm0.07^{\rm e}$\\
NGC 4258 &$-2.60\pm0.18$&$-2.52\pm0.17$&\nodata\\
NGC 4394 &$-2.42\pm0.32$&$-2.46\pm0.35$&\nodata\\
NGC 4395 &$-1.88\pm0.29$&$-1.95\pm0.23$&$-1.70\pm0.10^{\rm c}$\\
NGC 4736 &$-2.38\pm0.21$&$-2.30\pm0.11$&\nodata\\
NGC 5055 &$-2.74\pm0.32$&$-2.40\pm0.32$&\nodata\\
NGC 5236 (field 1) &$-2.40\pm0.09$&$-2.32\pm0.08$&$-2.25\pm0.12^{\rm b}$,\\
&&& $-2.38\pm 0.11^{\rm c}$\\
NGC 5236 (field 2) &$-2.73\pm0.26$&$-2.64\pm0.29$&\nodata\\
NGC 5457 (field 1) &$-2.75\pm0.13$&$-2.58\pm0.14$&\nodata\\
NGC 5457 (field 2) &$-2.45\pm0.11$&$-2.41\pm0.12$&\nodata\\
NGC 6217 &$-2.59\pm0.21$&$-2.27\pm0.22$&\nodata\\
NGC 6503 &$-2.68\pm0.41$&$-2.42\pm0.55$&\nodata\\
NGC 7793 &$-2.34\pm0.18$&$-2.33\pm0.40$&$-1.99\pm0.13^{\rm c}$ \\[6pt]
Composite Galaxies\\[6pt]
SFR1$^{\rm f}$&$-2.89\pm0.22$&\nodata&\nodata\\
SFR2$^{\rm f}$&$-2.56\pm0.09$&\nodata&\nodata\\
SFR3$^{\rm f}$&$-1.95\pm0.02$&\nodata&\nodata\\
HT1$^{\rm g}$&$-2.30\pm0.05$&\nodata&\nodata\\*
HT2$^{\rm g}$&$-2.43\pm0.03$&\nodata&\nodata\\
HT3$^{\rm g}$\hphantom{5457 (field 2)}&$-2.17\pm0.30$&\nodata&\nodata\\
MB1$^{\rm h}$&$-2.52\pm0.13$&\nodata&\nodata\\
MB2$^{\rm h}$&$-2.41\pm0.07$&\nodata&\nodata\\
MB3$^{\rm h}$&$-2.36\pm0.04$&\nodata&\nodata\\
\enddata
\tablenotetext{a}{ Mora et al.\ 2007 -- (B mag)}
\tablenotetext{b}{{}Larsen 2002 -- (V mag)}
\tablenotetext{c}{  Mora et al.\ 2009 -- (V mag) }
\tablenotetext{d}{ Dolphin \& Kennicutt 2002 -- V mag -- NOTE: The original paper quotes a power law slope of $-1.50$ for a fit to the function $dN/d\log L$ rather than the more standard $dN/dL$ used in this paper. A correction of $-1$ has therefore been  added to their value to make the comparison.} 
\tablenotetext{e}{ Whitmore et al.\ 2010 -- (V mag)}
\tablenotetext{f}{Galaxies Included --  SFR1 = 4258, 4394, 4736, 5055, 6503, 45, 1483, 4395, 5236 (field~2), 7793; SFR2 = 1313, 406, 628, 5236 (field~1), 5457 (field~1), 5457 (field~2); SFR3 = 1300 (field~1), 1300 (field~2), 3627, 4038, 1309}
\tablenotetext{g}{Galaxies Included -- HT1 = 4736, 4394, 3627, 2397; HT2 = 406, 5236 (field~1), 5236 (field~2), 628, 1483, 4258, 5055, 1309, 1300 (field~1), 1300 (field~2), 6217, 5457 (field~1), 5457 (field~2), 6503, 7793; HT3 = 1313, 45, 4395}
\tablenotetext{h}{Galaxies Included -- MB1 = 1483, 4395, 6503, 7793, 1313, 406, 4394, 4736; MB2 = 628, 6217, 5055, 1300 (field~1), 1300 (field~2), 4258; MB3 = 
2397, 3627, 5457 (field~1), 5457 (field~2), 5236 (field~1), 5236 (field~2)}
\end{deluxetable}
\end{document}